\documentclass[a4paper]{article}

\usepackage[english]{babel}
\usepackage[utf8x]{inputenc}
\usepackage[T1]{fontenc}
\usepackage{braket}

\usepackage[a4paper,top=3cm,bottom=2cm,left=3cm,right=3cm,marginparwidth=1.75cm]{geometry}

\usepackage{amsmath}
\usepackage{graphicx}
\usepackage[colorinlistoftodos]{todonotes}
\usepackage[colorlinks=true, allcolors=blue]{hyperref}
\usepackage{bm} 
\usepackage{epsfig}
\usepackage{titlesec}
\usepackage{dcolumn}
\usepackage[toc,page]{appendix}

\usepackage{xcolor}

\titleformat{\paragraph}
{\normalfont\normalsize\bfseries}{\theparagraph}{1em}{}
\titlespacing*{\paragraph}
{0pt}{3.25ex plus 1ex minus .2ex}{1.5ex plus .2ex}

\date{}

\title{\bf Constraints on  $\bm{2\ell 2q} $ \bf operators from  $\bm{\mu \leftrightarrow e}$
flavour-changing meson decays  }
\author{Sacha Davidson$^2$, Albert Saporta$^1$\\ \\
  \emph{$^1$  Universit\'e de Lyon,  France;  Universit\'{e}  Lyon 1,}\\
  \emph{ CNRS/IN2P3   IPNL,  69622 Villeurbanne cedex,
    France}\\ \\
\emph{$^2$ LUPM, CNRS,
Universit\'e Montpellier,}\\
\emph{Place Eug\`ene Bataillon, F-34095 Montpellier, Cedex 5, France}}

\begin{document}


\maketitle

\begin{abstract}
  We study  lepton flavour violating two- and three-body decays of
  pseudoscalar mesons in Effective Field Theory (EFT).  We give analytic formulae for the decay rates in the presence of a complete basis of  QED$\times$QCD-invariant operators.  The constraints are obtained at the experimental scale, then translated to the weak scale via  one-loop RGEs. The large RG-mixing between tensor and (pseudo)scalar operators
 weakens the  constraints on scalar and pseudoscalar operators at the weak scale.
\end{abstract}

\section{Introduction}

\noindent
    {
      The discovery of neutrino oscillations \cite{numassesoscill,neutrinoosci2}
established non zero neutrino masses  and  mixing angles \cite{PDG}.
If 
neutrinos are taken massless in the Standard Model (SM),
then New Physics (NP) is required to explain
the oscillation data. 
There are  several possibilities to search for NP signatures,
such as looking for new particles 
at the  LHC \cite{LHC1,LHC2}. Another possibility
is to look  for new processes among
known SM particles, such as Charged Lepton Flavour Violation
(CLFV) \cite{lowLFV3,CLFVintrol}, which we define  to be a contact
interaction that changes the flavour of charged leptons.
If neutrinos have renormalizable masses via  the Higgs mechanism,
then their contribution to CLFV rates is GIM  suppressed by a factor $\propto (m_{\nu}/M_W)^4 \sim 10^{-48}$.  
However, various  extensions of the
Standard Model that contain heavy new particles
(see {\it e.g.} \cite{lowLFV3,CLFVintrol,Arganda:2004bz,Raidal:2008jk}
and references therein),
can predict CLFV rates  comparable to the current
experimental sensitivities. Indeed,
low energy precision experiments searching for  forbidden SM modes, are sensitive to NP scales ≫ TeV \cite{lowLFV3}. Many
experiments search for  CLFV processes;  for example,
$\mu \leftrightarrow e$  flavour change can be probed in the decays $\mu \rightarrow e \gamma$ \cite{MEG} and $\mu \rightarrow 3e $ \cite{SINDRUM,mu3e}, in $\mu \rightarrow e$ conversion  on nuclei \cite{SINDRUM2,Comet,Mu2e} and also in meson decays such as $K,D,B \rightarrow \overline{\mu} e$ \cite{PDG,K0L,D0,B0S,K+e+,D+,Bpi+,BK+}. \\} In this paper, we focus on  leptonic and semileptonic pseudoscalar meson decays with a $\mu^{\pm} e^{\mp}$ in the final state \cite{PDG}.
    We assume  that these decays  could be mediated by  two-lepton, two-quark contact interactions, induced  by  heavy New Particles at the scale
    $\Lambda_{NP} > m_W$. The  contact interactions are included in a
    bottom-up
 Effective Field Theory (EFT) \cite{EFTgeorgi1,Manohar:2018aog,Pich:2018ltt} approach, 
 as a complete set of dimension six,  QED$\times$QCD-invariant operators \cite{lowLFV3},
 containing   a muon, an electron and one of the  quark-flavour-changing combinations $ds$, $bs$, $bd$ or $cu$.

    \noindent Many studies on related topics can be found in the literature.
    The experimental sensitivity to the coefficients of four-fermion operators (sometimes refered to as one-operator-at-a-time bounds), evaluated at the experimental scale, has been compiled by various authors \cite{LQ,tau,Carpentier}.
  Reference \cite{Cai:2015poa}   compared the sensitivities of the LHC vs
     low-energy processes, to quark flavour-diagonal scalar operators.
    The constraints  on combinations of   lepton-flavour-changing operator coefficients,
 which can be obtained from   the decays of same-flavour mesons,
 were studied in \cite{Petrov16}, and  the  radiative
 decays of $B,D$ and $K$ mesons were discussed in \cite{Petrov17}. 
        Lepton flavour-conserving, but quark flavour-changing  meson  decays (which occur in the Standard Model), are widely studied \cite{Buras}.
    In particular,  $B$ decays attract much current interest, due to the  observed
    anomalies
    \cite{LHCb1,LHCb2,LHCb,belle,babar}  which 
    suggest lepton universality violation \cite{SDG,Luca,SJ14,SJ17,Gundrun1,Bobeth,APS}.
 Lepton flavour change has been widely studied in various
  models (see {\it e.g.} references of \cite{lowLFV3,CLFVintrol,MLM}).     
  More model-independent studies, that take into account loop corrections
  (or equivalently, renormalization group running) have also been performed
 for $\mu \leftrightarrow e$ flavour change  \cite{RGE1,RGE2}.
Finally, with respect to the calculations in
this manuscript, the leptonic  branching ratio of
pseudoscalar mesons  is well-known,
and  can be found in \cite{LQ,Carpentier,Shanker,Bonn} and 
semi-leptonic branching ratios in various scenarios  can be found in \cite{Khadroelemevol,Dhadroelemevol,Bhadroelemevol,semil1,semil2,semil3,semil4,Crivellin:2015era,Crivellin:2016vjc}.

   The aim of this paper is to obtain constraints on the operator coefficients describing  meson decays at the experimental scale, and then transport the bounds
 to the weak scale\footnote{In a future publication,  
   we will give the evolution from the weak scale  to the NP scale, and discuss the prospects for
   reconstructing   the fundamental Lagrangian of the New Physics from
   the operator coefficients.}.
 The four fermion operators  that could induce the meson decays
 are listed in section \ref{sec:2}.  Section \ref{sec:3} 
 gives  the branching ratios for the leptonic and semileptonic  decays as a function of the operator coefficients. In section \ref{sec:4}, we then use the available bounds to constrain  the coefficients at the experimental scale ($\Lambda_{exp}\sim2$ GeV)
 by computing a covariance matrix, which allow us to take into account
 the interferences among the operators.
 The bounds are then evolved from the experimental scale to
 the weak scale ($\Lambda_{W}\sim m_W$)  in section \ref{sec:5}, using
 the Renormalization Group Equations (RGEs)   of QED and QCD for
 four-fermion operators  \cite{RGE1,RGE2}.
As discussed in the final section, these equations give a significant mixing of tensor operators into the (pseudo)scalars between $\Lambda_{exp}$ and $\Lambda_{W}$, which significantly weakens the bounds on (pseudo)scalar coefficients at $\Lambda_{W}$. \\

\section{A basis of   $\mu-e$ interactions  at low energy}
\label{sec:2}

We are interested in four-fermion operators
involving an electron, a muon
and two  quark of different  flavours,
which   are constructed  with chiral fermions, because
the lepton masses are frequently neglected, and
it simplifies the matching at the weak scale onto
SU(2)-invariant operators.  The operators
are added to the Lagrangian as
\begin{equation}
\delta {\cal L} = + 2\sqrt{2} G_F \sum_{O} \sum_{\zeta}{
C _O^{\zeta}} \mathcal{O}_O^{\zeta}  + h.c. 
\label{L}
\end{equation}

\noindent where  the subscript O identifies the Lorentz structure,
the superscript $\zeta=l_1 l_2 q_i q_j $
gives the flavour indices, and both run over the
possibilities in the lists below, extrapolated from
  \cite{lowLFV3,KKO}: 
  
  \begin{equation}
  \begin{array}{llll}
&\mathcal{O}^{e\mu uc}_{V,YY} =(\overline{e} \gamma^{\alpha} P_Y \mu ) 
(\overline{u} \gamma_{\alpha} P_Y c ), &&
\mathcal{O}^{e\mu uc}_{V,YX} = (\overline{e} \gamma^{\alpha} P_Y \mu ) 
(\overline{u} \gamma_{\alpha} P_X c )   \\
&\mathcal{O}^{e\mu cu}_{V,YY} = (\overline{e} \gamma^{\alpha} P_Y \mu ) 
(\overline{c} \gamma_{\alpha} P_Y u ), &&
\mathcal{O}^{e\mu cu}_{V,YX} = (\overline{e} \gamma^{\alpha} P_Y \mu ) 
(\overline{c} \gamma_{\alpha} P_X e)   \\
&\mathcal{O}^{e\mu uc}_{S,YY} = (\overline{e}  P_Y \mu ) 
(\overline{u} P_Y c ), &&
\mathcal{O}^{e\mu uc}_{S,YX} = (\overline{e}  P_Y \mu ) 
(\overline{u} P_X c )  \\
&\mathcal{O}^{e\mu cu}_{S,YY} = (\overline{e}  P_Y \mu ) 
(\overline{c} P_Y u ) , &&
\mathcal{O}^{e\mu cu}_{S,YX} = (\overline{e}  P_Y \mu ) 
(\overline{c} P_X u )  \\
&\mathcal{O}^{e\mu uc}_{T,YY} = (\overline{e} \sigma P_Y \mu ) 
(\overline{u} \sigma P_Y c ) && \\
&\mathcal{O}^{e\mu cu}_{T,YY} = (\overline{e} \sigma P_Y \mu ) 
(\overline{c} \sigma P_Y u ) &&
\end{array}
\label{opD}
\end{equation}

\begin{equation}
\begin{array}{llll}
&\mathcal{O}^{e\mu ds}_{V,YY} = (\overline{e} \gamma^{\alpha} P_Y \mu ) 
(\overline{d} \gamma_{\alpha} P_Y s ),  &&
\mathcal{O}^{e\mu ds}_{V,YX} = (\overline{e} \gamma^{\alpha} P_Y \mu ) 
(\overline{d} \gamma_{\alpha} P_X s )  \\
&\mathcal{O}^{e\mu sd}_{V,YY} = (\overline{e} \gamma^{\alpha} P_Y \mu ) 
(\overline{s} \gamma_{\alpha} P_Y d ),  &&
\mathcal{O}^{e\mu sd}_{V,YX} = (\overline{e} \gamma^{\alpha} P_Y \mu ) 
(\overline{s} \gamma_{\alpha} P_X d )   \\
&\mathcal{O}^{e\mu ds}_{S,YY} = (\overline{e}  P_Y \mu ) 
(\overline{d} P_Y s ),  &&
\mathcal{O}^{e\mu ds}_{S,YX} = (\overline{e}  P_Y \mu ) 
(\overline{d} P_X s )   \\
&\mathcal{O}^{e\mu sd}_{S,YY} = (\overline{e}  P_Y \mu ) 
(\overline{s} P_Y d ),  &&
\mathcal{O}^{e\mu ds}_{S,YX} = (\overline{e}  P_Y \mu ) 
(\overline{d} P_X s )  \\
&\mathcal{O}^{e\mu ds}_{T,YY} = (\overline{e} \sigma P_Y \mu ) 
(\overline{d} \sigma P_Y s ) &&   \\
&\mathcal{O}^{e\mu sd}_{T,YY} = (\overline{e} \sigma P_Y \mu ) 
(\overline{s} \sigma P_Y d ) &&   \\
\end{array}
\label{opKaon}
 \end{equation}
 
\noindent where
$YY \in \{LL,RR\}$, and  $X Y  \in \{LR,RL\}$, and
the list is given explicitly for the Kaon and $D$-meson operators.
The lists for   the $B_d$  and $B_s$ are
obtained from eqn. \eqref{opKaon} by replacing
$ds \to db,sb$.
The operators
  are normalised such that the Feynman
rule will be $+i C/\Lambda^2$. 
The operators in the lists \eqref{opD} and \eqref{opKaon} 
transform a muon into an electron; the $e \to \mu$
operators arise  in the $+h.c.$ of eqn (\ref{L}). So in
these conventions, the lepton flavour indices
are always $e\mu$, and  do not need to be
given. 
In the
following sections,  we  give the  decay rates of
pseudoscalar mesons, composed of constituent quarks
$\bar{q}_i $ and $q_j$, into  $e^+ \mu^-$ or $e^- \mu^+$. Then
we obtain constraints on the operator coefficients by
comparing to the experimental upper bounds on
the branching ratios, {\it e.g.}
$
BR(P_1 \to e^\pm \mu^\mp) =
BR(P_1 \to e^+ \mu^-) +
BR(P_1 \to e^- \mu^+) < ...
$
which we suppose to apply independently to
both decays. This gives  independent and identical bounds
on $\epsilon^{e\mu q_iq_j}$ and  $\epsilon^{e\mu q_jq_i}$.

\noindent In this work, we choose an operator basis  with non-chiral quark currents, which is convenient for the non-chiral hadronic matrix elements involved in meson decays.  Thus, the operators describing the contact interactions that can mediate leptonic ($\overline{q_i}q_j \rightarrow \overline{\mu}e $) and semileptonic ($q_i \rightarrow q_j\overline{\mu}e $) CLFV pseudoscalar meson decays  at a scale $\Lambda_{exp} \sim 2$ GeV  ($\Lambda_{exp}\sim m_b \simeq  4.2$ GeV for $b$s and $b$d) are written: 
 \begin{equation}
\begin{split}
&\mathcal{O}^{ e \mu q_i q_j }_{S,X} = \left( \overline{e}  P_{X} \mu \right) 
\left( \overline{q_i }   q_j \right), \quad \mathcal{O}^{e\mu q_i q_j }_{P,X} = \left( \overline{e}P_{X} \mu \right) 
\left( \overline{q_i }  \gamma^{5}   q_j \right) \\
&\mathcal{O}^{e\mu q_i q_j }_{V,X} = \left( \overline{e} \gamma^{\alpha} P_{X} \mu \right) 
\left( \overline{q_i} \gamma_{\alpha} q_j \right), \quad \mathcal{O}^{e\mu q_i q_j }_{A,X} = \left( \overline{e} \gamma^{\alpha} P_{X} \mu \right) 
\left( \overline{q_i} \gamma_{\alpha}\gamma^{5}  q_j \right)  \\
&\mathcal{O}^{e \mu q_i q_j }_{T_X} = \left( \overline{e} \sigma^{\alpha \beta} P_{X} \mu \right) \left( \overline{q_i } \sigma_{\alpha \beta}P_X  q_j \right)
\end{split}
\label{operators}
\end{equation} \\

\noindent where  $q_{i,j} \in \left\{u,d,s,c,b\right\},$ $P_{X}= P_{R,L} = \frac{1 \pm \gamma_5}{2}$ and $\sigma^{\mu \nu} = \frac{i}{2}[\gamma^{\mu},\gamma^{\nu}]$. \\

\noindent 
In this case, the coefficients $\epsilon$ of the operators  in eqn.  \eqref{operators}   are 
 :

\begin{equation}
\begin{split}
&\epsilon_{S,X}^{e\mu q_i q_j } =\frac{1}{2}(C_{S,XR}^{e\mu  q_i q_j }+C_{S,XL}^{e\mu  q_i q_j }) , \quad 
\epsilon_{P,X}^{e\mu  q_i q_j } =\frac{1}{2}(C_{S,XR}^{e\mu q_i q_j }-C_{S,XL}^{e\mu  q_i q_j }) \\
&\epsilon_{V,X}^{e\mu  q_i q_j } =\frac{1}{2}(C_{V,XR}^{e\mu  q_i q_j }+C_{V,XL}^{e\mu  q_i q_j }) , \quad
\epsilon_{A,X}^{e\mu  q_i q_j } =\frac{1}{2}(C_{V,XR}^{e\mu  q_i q_j }-C_{V,XL}^{e\mu  q_i q_j }) \\
&\epsilon_{T,X}^{e\mu  q_i q_j } =C_{T,XX}^{e\mu  q_i q_j } 
\end{split}
\label{coefficients}
\end{equation}

\noindent In the next section, we compute the branching ratio for the (semi)leptonic decays as a function of  the coefficients of eqn. \eqref{coefficients}.

\section{Leptonic and semileptonic pseudoscalar meson decays}
\label{sec:3}

\noindent There are a multitude of bounds   on rare meson decays coming from precision experiments \cite{PDG,Carpentier}. The aim of this paper is to use these bounds to constrain the coefficients of eqn. \eqref{coefficients}. Thus, in this section, we compute the leptonic and semileptonic pseudoscalar meson decay branching ratio as a function of these coefficients.

\subsection{Leptonic decay branching ratio}
\noindent We are interested in decays such as :
$P_1 \rightarrow l_1 \bar{l}_2  $  where $\left\{l_1, l_2 \right\}$  are leptons of mass $m_1,m_2$ and $P_1$ is a pseudoscalar meson of mass M ($P_1 \in \left\{K^0_L(\frac{\bar{d}s+\bar{s}d}{\sqrt{2}}),D^0(\bar{u}c),B^0(\bar{b}d) \right\}$).  
\noindent In the presence of New Physics, the leptonic decay branching ratio of a pseudoscalar meson $P_1$ of mass $M$ is written \cite{LQ,Carpentier,Bonn}:

 \begin{equation}
\begin{split}
\frac{BR(P_1 \rightarrow l_1 \bar{l}_2  )}{C_{2\text{body}}}&=
(|\epsilon_{P,L}|^2+|\epsilon_{P,R}|^2) \tilde{P'}^2 (M^2 -m_1^2-m_2^2) \\
&+ (|\epsilon_{A,L}|^2+|\epsilon_{A,R}|^2) \tilde{A'}^2[(M^2-m_1^2-m_2^2)(m_1^2+m_2^2) +4m_1^2m_2^2] \\
&- 2(\epsilon_{P,L}\epsilon_{A,R} + \epsilon_{P,R}\epsilon_{A,L})\tilde{P'}\tilde{A'}
m_2(M^2 +m_1^2 -m_2^2)  \\
&+ 2(\epsilon_{P,L}\epsilon_{A,L} + \epsilon_{P,R}\epsilon_{A,R})\tilde{P'}\tilde{A'}
m_1(M^2 +m_2^2 -m_1^2) \\
&-4\epsilon_{P,L}\epsilon_{P,R}\tilde{P'}^2
m_1m_2 \\
&-4\epsilon_{A,L}\epsilon_{A,R}\tilde{A'}^2
M^2m_1m_2  \\
\end{split}
\label{leptonic rate}
\end{equation} \\

\noindent where $C_{2\text{body}} =\frac{\tau r^*G_F^2}{\pi M^2}$, $ r^*= \frac{1}{2M}\sqrt{(M^2-(m_1+m_2)^2)(M^2-(m_1-m_2)^2)} $, $m_{1,2}$ are the masses of the leptons and $\tau$ is the lifetime of $P_1$. For simplicity, we dropped the flavour superscript ($\zeta=l_1 l_2 q_i q_j$) of the coefficients. \\

\noindent The expectation values of the quark current for a pseudoscalar meson  are written \cite{Carpentier,Bonn} : 

\begin{equation}
\tilde{P'}=\frac{1}{2}\braket{0|\bar{q_i} \gamma^{5} q_j|P_1} =\frac{f_{P_1} M^2}{2(m_{i} +m_{j})}, \quad\tilde{A'k^{\mu}}=\frac{1}{2}\braket{0|\bar{q_i}\gamma^{\mu} \gamma^{5} q_j|P_1}=\frac{f_{P_1}k^{\mu}}{2}
\label{expt value}
\end{equation}

\noindent where $m_{i,j}$ are the masses of the quarks, $f_{P_1}$  is the decay constant of the meson and $k^{\mu}$  the momentum of the meson. 
These formulae are used for pions, Kaons, D and B mesons. The values of the constants are given in appendix \ref{appendix:Constants}. Note that tensor operators do not contribute to the leptonic decay, because the trace of product of the Dirac matrices contained in the tensor operator vanishes in this case.

\subsection{Semileptonic decay branching ratio}


\noindent We are interested in decays such as :
$P_1 \rightarrow l_1 \bar{l}_2   P_2$  where $\left\{l_1, l_2 \right\}$  are leptons of mass $m_1,m_2$ and $\left\{P_1,P_2\right\}$ are pseudoscalar mesons of mass $M,m_3$ ($P_1 \in \left\{K^+(u\bar{s}),D^+(c\bar{d}),B^+(u\bar{b}),B_s^+(s\bar{b}) \right\}$ and $P_2 \in \left\{\pi^+(u\bar{d}),K^+(u\bar{s}) \right\}$). The semileptonic decay branching ratio is written \cite{Ilisie}:
\begin{equation}
\hspace{-0.8cm}
BR(P_1 \rightarrow l_1 \bar{l}_2   P_2)=\frac{\tau}{512\pi^3M^3}\frac{1}{2J+1}\int_{(m_1+m_2)^2}^{(M-m_3)^2}dq^2 \int_{-1}^{1}d\cos\theta \frac{|\mathcal{M}|^2\sqrt{\lambda(M^2,m_3^2,q^2)}\sqrt{\lambda(q^2,m_1^2,m_2^2)}}{q^2} 
\label{3 body rate phase space}
\end{equation}

\noindent where $q=(p_1 + p_2)$ is the transferred momentum, $\theta$ the angle between the direction of propagation of the lighter meson ($P_2$) and the antilepton ($l_2$) in the leptons reference frame, $\tau$ and J  the lifetime and the spin  of $P_1$  and $|\mathcal{M}|^2$ the matrix element of the semileptonic decay. The Kallen function is defined as $\lambda(x,y,z) = (x-y-z)^2 - 4yz$.\\

\noindent In the presence of New Physics, the matrix element in the semileptonic decay branching ratio of eqn. \eqref{3 body rate phase space} is written :

\begin{equation}
\begin{split}
\frac{|\mathcal{M}|^2}{8 G_F^2} &= 2(|\epsilon_{S,L}|^2+|\epsilon_{S,R}|^2) \tilde{S}^2 (p_1.p_2) \\[10pt]
& + \frac{1}{4}(|\epsilon_{V,L}|^2 + |\epsilon_{V,R}|^2 ) [f_+^2 \left(
 4(p_1.P)(p_2.P) - 2P^2 (p_1.p_2) \right) + f_-^2 \left(
 4(p_1.q)(p_2.q) - 2q^2 (p_1.p_2) \right)\\
 & + 4f_+f_-   \left((p_1.q)(p_2.P) + (p_1.P)(p_2.q)-(p_1.p_2)(P.q)    \right)] \\[10pt]
 \vspace{0.4cm}
 & + 4(|\epsilon_{T_{R}}|^2 + |\epsilon_{T_L}|^2 ) \tilde{T'}^2 [ 4 (p_1.q)(p_2.P)(P.q) + 4(p_1.P)(p_2.q)(P.q) -2(p_1.p_2)(P.q)^2  \\
 &+2P^2q^2(p_1.p_2) - 4 P^2(p_1.q)(p_2.q) - 4q^2(p_1.P)(p_2.P) 
 ] \\[10pt]
 &-2(\epsilon_{S,L}\epsilon_{V,R}+\epsilon_{S,R}\epsilon_{V,L})\tilde{S}m_2[ \left(f_+ (p_1.P) + f_-(p_1.q) \right) ] \\[10pt]
  &+2(\epsilon_{S,L}\epsilon_{V,L}+\epsilon_{S,R}\epsilon_{V,R})\tilde{S}m_1[ \left(f_+ (p_2.P) + f_-(p_2.q) \right) ] \\[10pt]
 & +8(\epsilon_{S,R}\epsilon_{T_R}+\epsilon_{S,L}\epsilon_{T_L})\tilde{S}\tilde{T'}[\left((p_1.P)(p_2.q)-(p_1.q)(p_2.P)  \right) ] \\[10pt]
  &-4\epsilon_{S,L}\epsilon_{S,R}\tilde{S}^2 m_1 m_2 \\[10pt]
   &-\epsilon_{V,L}\epsilon_{V,R} m_1 m_2[f_-^2q^2 + f_+^2P^2 + 2 f_+ f_-(P.q)]\\[10pt]
     &+4(\epsilon_{V,L}\epsilon_{T_{R}} +\epsilon_{V,R}\epsilon_{T_{L}})
     \tilde{T'}m_2[f_+((p_1.q)p^2-(P.p_1)(P.q)) +  f_-((p_1.q)(P.q)-(p_1.P)q^2)]\\[10pt]
      & +4 (\epsilon_{V,R}\epsilon_{T_R}+\epsilon_{V,L}\epsilon_{T_{L}})\tilde{T'}m_1[ \left(f_+((P^2)(p_2.q)-(p_2.P)(P.q)) +f_-((p_2.q)(P.q)-q^2(p_2.P))
 \right)] \\[10pt] 
 &+16\epsilon_{T_{R}}\epsilon_{T_{L}}\tilde{T'}^2 m_1 m_2[(P.q)^2-P^2q^2]\\
\end{split}
\label{M3body}
\end{equation} \\

\normalsize
\noindent where  $p_{1},p_{2},k,p_{3}$  are respectively  the 4-momentum of the leptons 1 and 2,  and  the 4-momenta of $P_1$ and $P_2$,    $P = k +p_3$ and  the hadronic matrix elements are  written \cite{Carpentier,Bonn,Khadroelemevol,Dhadroelemevol,Bhadroelemevol} : 
\begin{equation}
\begin{split}
&\tilde{V}^\mu =\frac{1}{2}\braket{P_2|\bar{q_i} \gamma^{\mu}q_j|P_1}=\frac{1}{2}(P^{\mu}f_+^{P_1 P_2}(q^2) + q^{\mu}f_-^{P_1 P_2}(q^2) ) \\
&\tilde{S} =\frac{1}{2} \braket{P_2|\bar{q_i}   q_j|P_1}=\frac{1}{2}\frac{(M^2-m_{3}^2)}{(m_{q_i}-m_{q_j})}f_0^{P_1 P_2}(q^2)\\
&\tilde{T}^{\mu \nu} = \frac{1}{2} \braket{P_2|\bar{q_i}\sigma^{\mu \nu}q_j|P_1} =-\frac{i}{2} \frac{(f_+^{P_1 P_2}(q^2)-f_-^{P_1 P_2}(q^2))}{M^*}(P^{\mu}q^{\nu}-P^{\nu}q^{\mu}) \\ 
&\tilde{T'} = \frac{1}{2} \frac{(f_+^{P_1 P_2}(q^2)-f_-^{P_1 P_2}(q^2))}{M^*}
\end{split}
\end{equation} 
For simplicity, we suppressed the $q^2$ dependence of the form factors $f_{+,-,0}$ in eqn. \eqref{M3body},  and   the
 flavour  superscript ($\zeta=l_1 l_2 q_i q_j $) of the coefficients.
  Notice there is no interference between $\epsilon_{S,L}$ ($\epsilon_{S,R}$ ) and $\epsilon_{T_R}$ ($\epsilon_{T_L}$) because the  trace of the product of Dirac matrices involved in tensor and scalar operators of different chirality vanishes.  The form factors and the scalar product in eqn. \eqref{M3body} are given in appendix \ref{appendix:kin-formfactor}. 

\noindent For simplicity, we do not give the analytic expression of the integrated semileptonic decay branching ratio, but only perform the integrals numerically.

\section{Covariance matrix}
\label{sec:4}

\noindent In this section, we use  the Branching Ratios (BRs) of
  eqns \eqref{leptonic rate} and \eqref{3 body rate phase space} to compute a covariance matrix, that will give constraints on  the coefficients that account
   for possible interferences. We note $BR_2^{exp}$ [$BR_3^{exp}$] the experimental upper limit on the leptonic decay $P_1 \rightarrow \bar{l}_1l_2$ [semileptonic decay $P_1 \rightarrow P_2\bar{l}_1l_2$] branching ratio and $M_2$ [$M_3$] the associated covariance matrix.

\noindent We can write the decay branching ratio of eqn. \eqref{leptonic rate}  and \eqref{3 body rate phase space} in the form 
\begin{equation}
\vec{\epsilon}^TM^{-1}\vec{\epsilon} =1
\label{CMmpoins1C}
\end{equation}

\noindent where  $\vec{\epsilon}^T (\vec{\epsilon})$ is a row (column)
  vector of coefficients, and $M^{-1}$ is the inverse of the covariance matrix. The explicit form of the $4 \times 4$ and $6 \times 6$ matrices is given in appendix \ref{appendix:M2/M3-1}. The diagonal elements  of the covariance matrix $M$
represent the squared bounds on our coefficients, and the off-diagonals elements represent the correlation between coefficients. 

\begin{center}
\renewcommand{\arraystretch}{2}
\begin{table}[!htb]
\hspace{0.4cm}
\small
\begin{tabular}{|c|c|c|c|c|c|c|c|c|}
  \hline
  $Decay$ &$Leptonic$&$Semileptonic$  \\
    \hline
   $K$ &$BR_2^{exp}(K^0_L \rightarrow \mu^{\pm}e^{\mp})<4.7 \times 10^{-12} \quad \cite{K0L}$&$BR_3^{exp}(K^+ \rightarrow \pi^+ \bar{\mu}e) < 1.3 \times 10^{-11}$ \\
&-&$BR_3^{exp}(K^+ \rightarrow \pi^+ \bar{e}\mu) < 5.2 \times 10^{-10}  \quad \cite{K+e+}$\\
  \hline
     $D $ &$BR_2^{exp}(D^0 \rightarrow \mu^{\pm} e^{\mp})<1.3 \times 10^{-8} \quad \cite{D0}$&$BR_3^{exp}(D^+ \rightarrow \pi^+ \bar{\mu}e) < 3.6 \times 10^{-6}$\\
  &-&$BR_3^{exp}(D^+ \rightarrow \pi^+ \bar{e}\mu) < 2.9 \times 10^{-6} \quad \cite{D+}$\\   
  \hline
  $D_s $ &-&$BR_3^{exp}(D^+_S \rightarrow K^+ \bar{\mu}e) < 9.7 \times 10^{-6}
$\\
  &-&$BR_3^{exp}(D^+_S \rightarrow K^+ \bar{e}\mu) < 1.4 \times 10^{-5} \quad \cite{D+}$\\   
  \hline
     $B $ &$BR_2^{exp}(B^0 \rightarrow \mu^{\pm}e^{\mp})<2.8 \times 10^{-9} \quad \cite{B0S}$&$BR_3^{exp}(B^+ \rightarrow \pi^+ \mu^{\pm}e^{\mp}) < 1.7 \times 10^{-7} \quad \cite{Bpi+}$\\
     &-&$BR_3^{exp}(B^+ \rightarrow K^+ \mu^{\pm}e^{\mp}) < 9.1 \times 10^{-8}  \quad \cite{BK+}$\\
  \hline
       $B_s$ &$BR_2^{exp}(B^0_S \rightarrow \mu^{\pm}e^{\mp})<1.1 \times 10^{-8}  \quad \cite{B0S}$&$-$\\
  \hline
\end{tabular}
   \caption{Experimental bounds on leptonic and semileptonic decays.}
\label{table:exptal bounds}
\end{table}
\renewcommand{\arraystretch}{1}
\end{center}

\newpage
\subsection{Bounds on the coefficients}

\noindent In this section, we give constraints on the coefficients for Kaon, D and B meson leptonic and semileptonic decays. \noindent As explained in section \ref{sec:3}, tensor operators do not contribute to the leptonic decays of mesons. Thus, the available upper limits on leptonic [semileptonic] pseudoscalar meson  branching ratios will give constrains on the $\epsilon_{P,X}$ and  $\epsilon_{A,X}$  [$\epsilon_{S,X}$, $\epsilon_{V,X}$ and  $\epsilon_{T,X}$] coefficients. Indeed,  hadronic matrix elements with  scalar, vector or tensor quark current structure vanish in the leptonic case, while hadronic matrix elements with pseudoscalar or axial struture vanish in the semileptonic case. We consider the  CLFV decays with the associated experimental upper limits given in table \ref{table:exptal bounds}
\cite{PDG}.

\noindent The bounds in table \ref{table:exptal bounds} will be used to constrain the coefficients at $\Lambda_{exp}$ and the at $\Lambda_W$ after the RGEs evolution of the coefficients (see section \ref{sec:5}). The covariance matrices at $\Lambda_{exp}$ for the (semi)leptonic meson decays are given in appendix \ref{appendix:bounds exp/w}, and the bounds on coefficients are summarized in table \ref{table:Scalar}, \ref{table:V} and \ref{table:tensor}.

\section{Renormalization Group Equations (RGEs)}
\label{sec:5}

\noindent In this section, we review the evolution  of operator coefficients from the experimental scale ($\Lambda_{exp}\sim$ 2 GeV) up to the weak scale ($\Lambda_{W}\sim$ 80 GeV) via the one-loop RGEs of QED and QCD \cite{RGE1,RGE2}. We only consider the QED$\times$QCD invariant operators of eqn. \eqref{operators}. The matching onto the SMEFT basis \cite{SMEFT2} and the running  above $m_W$ \cite{SMEFT3} will be studied  at a later date.

\subsection{Anomalous dimensions for meson decays}

\begin{figure}[h]
\begin{center}
\includegraphics[scale=0.9]{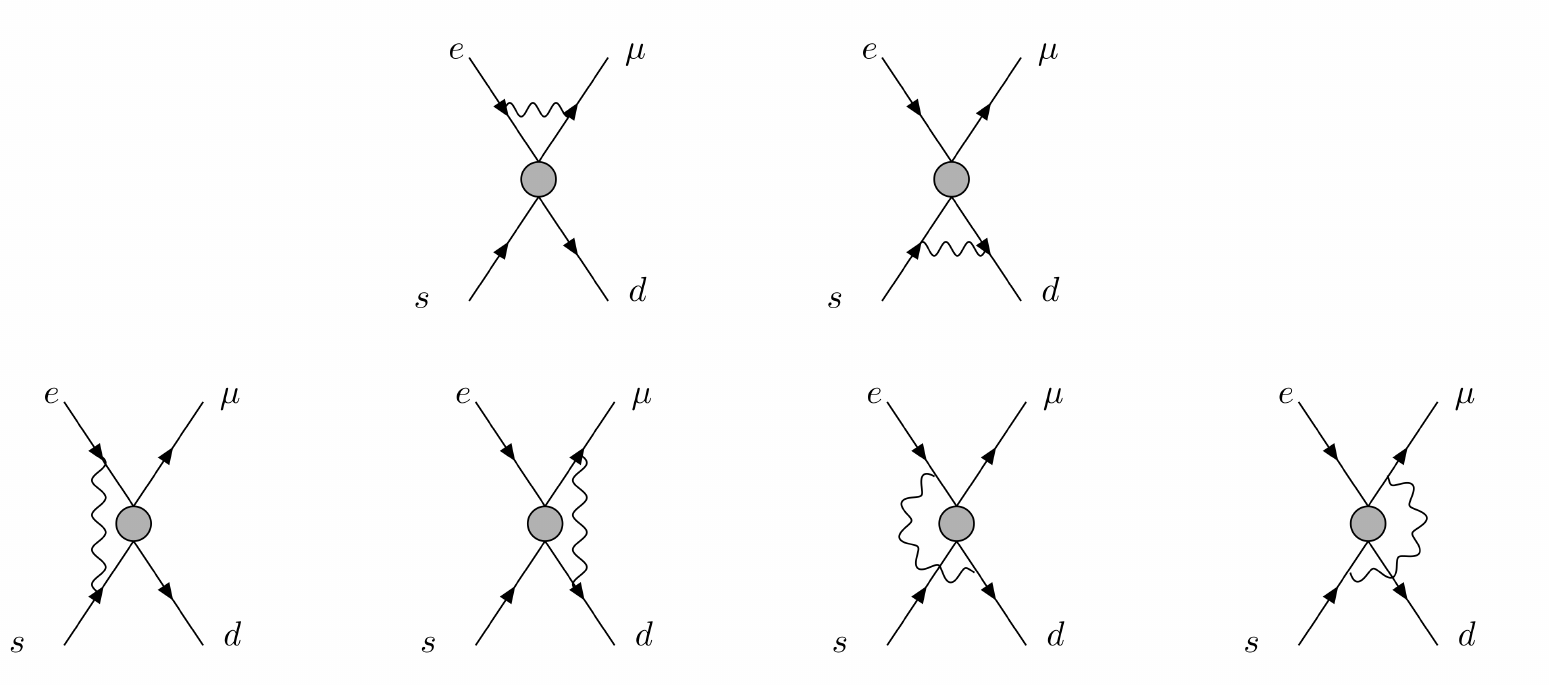}
\end{center}
\caption{
  Examples of one-loop gauge vertex corrections to 4-fermion operators.
The wave-function renormalization diagrams are missing.
 \label{fig:1 RG loops}
}
\end{figure}

Figure \ref{fig:1 RG loops} illustrates some of  the  one-loop
diagrams  that  renormalize  our  operators below the weak scale.
Operator mixing is  induced by photon loops, whereas the QCD
corrections only rescale the S,P and T operator coefficients.
After including one-loop corrections in the $\overline{MS}$ scheme, the operator coefficients will run with scale $\mu$ according to \cite{RGE1} :

\begin{equation}
 \mu\frac{\partial}{\partial \mu}\vec{\epsilon}= \frac{\alpha_e}{4\pi}\vec{\epsilon} \,\Gamma^e +\frac{\alpha_s}{4\pi}\vec{\epsilon} \,\Gamma^s
 \label{diffRGE}
\end{equation}

\noindent where  $\Gamma^e$ and $\Gamma^s$ are the QED and QCD anomalous dimension matrices  and $\vec{\epsilon}$ is a row vector that contains  the operator coefficients  of eqn. \eqref{coefficients}.  In this work, we use the approximate analytic solution   \cite{SachaKunoCiri} of eqn. \eqref{diffRGE} to compute the running and mixing of the coefficients between $\Lambda_{exp}$ and $\Lambda_{W}$ :

\begin{equation}
 \epsilon_{I} (\Lambda_{exp})  = \epsilon_{J}(\Lambda_{W})\lambda^{a_{J}}
\left( \delta_{JI} - \frac{\alpha_{e}  \widetilde{\Gamma}^{e}_{JI} }{4\pi}  \log \frac{\Lambda_{W}}{\Lambda_{exp}}  \right)
\label{RGE}
\end{equation} \\

\noindent where  I,J represent
the super- and sub-scripts which label operator
coefficients,  $\lambda$ encodes the
QCD corrections, and $\widetilde{\Gamma}^{e}_{JI}$  is the
``QCD-corrected''   one-loop, anomalous dimension matrix for QED \cite{FJI1,FJI2} . 
The elements of $\widetilde{\Gamma}^{e}_{JI}$ are defined as:
\begin{equation}
\widetilde{\Gamma}^{e}_{JI}=\Gamma^{e}_{JI} f_{JI}, \quad  f_{JI}=\frac{1}{1+a_{J} - a_{I}} \frac{ \lambda^{a_{I} - a_{J}} - \lambda}{1 - \lambda}
, \quad \Gamma^{e}=
\begin{bmatrix} 
\Gamma_{SPT} & 0 \\
0 & \Gamma_{VA} 
\end{bmatrix} ~~.
\label{gammaetilde}
\end{equation} \\
where there is no sum on $I,J$, 
$\lambda = \frac{ \alpha_{s}(\Lambda_{W})}{\alpha_{s}(\Lambda_{exp})}$,
 and $a_{J} = \frac{\Gamma_{JJ}^{s}}{2 \beta_{0}} = \left\{ -\frac{12}{23}, -\frac{12}{23}, \frac{4}{23} \right\}$ for $J \in \left\{S,P,T\right\}$. 
The QED anomalous dimensions are
\begin{align}
&\Gamma_{SPT}=
\left[
\begin{array}{cccccccccccc} 
& \gamma_{PP}^{l_1l_2q_iq_j} & 0 & \gamma_{PT}^{l_1l_2q_iq_j}   \\
&0 & \gamma_{SS}^{l_1l_2q_iq_j}  & \gamma_{ST}^{l_1l_2q_iq_j}  \\
&\gamma_{TP}^{l_1l_2q_iq_j}   &\gamma_{TS}^{l_1l_2q_iq_j} &\gamma_{TT}^{l_1l_2q_iq_j}   \\
\end{array}
\right]
, \quad \Gamma_{VA}=
\left[
\begin{array}{cccccccccccc} 
  & \gamma_{AA}^{l_1l_2q_iq_j} &\gamma_{AV}^{l_1l_2q_iq_j}  \\
 &\gamma_{VA}^{l_1l_2q_iq_j} & \gamma_{VV}^{l_1l_2q_iq_j}  \\
\end{array}
\right]
\end{align}

\noindent where the matrix elements in $\Gamma_{SPT}$
 and $\Gamma_{VA}$ 
are defined in section \ref{sec:5}. 



\newpage

For the scalars and pseudoscalars,
the  wavefunction, first and   second  diagrams  of figure \ref{fig:1 RG loops}
renormalize the  coefficients,
while the last four diagrams mix the tensors  into the scalars  and pseudoscalars:

\begin {equation}
  \gamma_ {SS}^{q, q} =
  \begin {array} {c | cc}
 &\epsilon_ {S, L}^{q q} &\epsilon_ {S, R}^{q q} \\\hline
 \epsilon_ {S, L}^{q q} & - 6 (1 + Q_q^2) & 0 \\
 \epsilon_ {S, R}^{q q} & 0 & - 6 (1 + Q_q^2) \\
 \end {array} ~~ ~
     \gamma_ {TS}^{q, q} =
  \begin {array} {c | cc}
 &\epsilon_ {S, L}^{q q} &\epsilon_ {S, R}^{q q} \\\hline
 \epsilon_ {T, L}^{q q} & 48 Q_q & 0 \\
 \epsilon_ {T, R}^{q q} & 0 & 48 Q_q \\
 \end {array}
 \label{ANomdimTS}~~~
\end {equation}
\begin {equation}
  \gamma_ {PP}^{q, q} =
  \begin {array} {c | cc}
 &\epsilon_ {P, L}^{q q} &\epsilon_ {P, R}^{q q} \\\hline
 \epsilon_ {P, L}^{q q} & -6 (1 + Q_q^2) & 0 \\
 \epsilon_ {P, R}^{q q} & 0 & -6 (1 + Q_q^2) \\
 \end {array} ~~ ~
     \gamma_ {TP}^{q, q} =
  \begin {array} {c | cc}
 & \epsilon_{P, L}^{q q} & \epsilon_{P, R}^{q q} \\\hline
 \epsilon_{T, L}^{q q} & - 48 Q_q & 0 \\
 \epsilon_{T, R}^{q q} & 0 & 48 Q_q \\
 \end {array}
  \label{ANomdimTP}~~~.
\end {equation} \\

\noindent Similarly, the last four diagrams mix  the (pseudo)scalars  to the  tensors. Only the wavefunction diagrams renormalize the tensors, because for the the first and second  diagrams $\gamma^{\mu} \sigma \gamma_{\mu}=0$. We obtain  :

\begin {equation}
 \gamma_ {TT}^{q, q} =
 \begin {array} {c | cc}
& \epsilon_{T, L}^{q q} & \epsilon_{T, R}^{q q} \\\hline
\epsilon_{T, L}^{q q} &  2 (1 + Q_q^2) & 0 \\
\epsilon_{T, R}^{q q} & 0 & 2 (1 + Q_q^2) \\
\end {array}
\hspace {0.5 cm}
\gamma_ {S(P)T}^{q, q} =
 \begin {array} {c | cc}
& \epsilon_{T, L}^{q q} & \epsilon_{T, R}^{q q} \\\hline
\epsilon_{S (P), L}^{q q} & (-)2 Q_q & 0 \\
\epsilon_{S (P), R}^{q q} & 0 & 2 Q_q \\
\end {array}
\end {equation}

Finally, for the vectors and axial vectors,
there is no running, but the last four diagrams
contribute to the mixing of vector and axial coefficients   :

\begin {equation}
 \gamma_ {AV}^{q, q} =
 \begin {array} {c | cc}
& \epsilon_{V, L}^{q q} & \epsilon_{V, R}^{q q} \\\hline
\epsilon_{A, L}^{q q}   &  12 Q_q &0 \\
\epsilon_{A, R}^{q q}  &0& - 12 Q_q \\
\end {array}
\hspace {1 cm}
 \gamma_ {VA}^{q, q} =
 \begin {array} {c | cc}
& \epsilon_{A, L}^{q q} & \epsilon_{A, R}^{q q} \\\hline
\epsilon_{V, L}^{q q} &  12 Q_q &0 \\
\epsilon_{V
, R}^{q q}  &0& - 12 Q_q \\
\end {array}
\end {equation}

\subsection{RGEs of operator coefficients}

\noindent In this section we compute the evolution of the bounds from $\Lambda_{exp}$ to   $\Lambda_{W}$. In  the previous section, we found a mixing between pseudoscalar and tensor coefficients, and between vector and axial coefficients. Thus, the coefficients that contributed only to the leptonic [semileptonic] decays at $\Lambda_{exp}$ will also contribute to the semileptonic [leptonic] decays at $\Lambda_{W}$ via the mixing. \\
The matrices describing the evolution of the coefficients from $\Lambda_{exp}$ to   $\Lambda_{W}$ for all the decays were obtained with eqn. \eqref{RGE} and are given in appendix \ref{appendix:RGESmatrices}.

\normalsize

\normalsize

\subsection{Evolution of the bounds}

  In order to constrain  the coefficients at $\Lambda_{W}$,
the constraints needs to be expressed in terms of  coefficients
at $\Lambda_{W}$. However, the mixing of the pseudoscalar (axial) into the tensor (vector),  and vice versa, implies that  leptonic and
semi-leptonic  branching ratios can both depend on any of the ten
 coefficients, which we arrange in a vector as
 $\vec{\epsilon'}= \left( \begin{array}{c}\epsilon_{P,L},\epsilon_{A,L},\epsilon_{P,R},\epsilon_{A,R},\epsilon_{S,L},\epsilon_{V,L},\epsilon_{T_{L}}, \epsilon_{S,R},\epsilon_{V,R},\epsilon_{T_{R}}\\\end{array} \right)_{\Lambda_W}$. The $10 \times 10$ matrix we need to invert to compute the bounds at $\Lambda_W$ is now written :

\begin{equation}
(M')^{-1}= \mathcal{R}^T
\left(
\begin{array}{cccccccccc}
 M_{2}^{-1} & 0_{4\times6}  \\
 0_{6\times4} & M_{3}^{-1}  \\
\end{array}
\right)\mathcal{R}
\label{M-1prime}
\end{equation}

\noindent where $M_{2}^{-1}$ and $M_{3}^{-1}$ are the $4\times4$  and $6\times6$ matrices defined in appendix \ref{appendix:M2/M3-1} we inverted to obtain the bounds at $\Lambda_{exp}$ (see section \ref{sec:4}) and $\mathcal{R}$ has the form of  the matrices defined in eqn.  \eqref{REGKaon}, \eqref{REGDmeson} and \eqref{REGBmeson}.  Finally, eqn. \eqref{CMmpoins1C}  is written in the new basis as : 

\begin{equation}
\vec{\epsilon '}^{T}(M')^{-1}\vec{\epsilon '} =1
\label{CMmpoins1CatMW}
\end{equation}

\noindent where $\vec{\epsilon '}$ is the vector of coefficients at $\Lambda_W$, $(M')^{-1}$ the matrix in eqn. \eqref{M-1prime}    and the superscript T  means matrix transposition. All the covariance matrices at $\Lambda_{W}$ can be found in appendix \ref{appendix:bounds exp/w}. In table \ref{table:Scalar}, \ref{table:V} and \ref{table:tensor} we summarize all the bounds on the coefficients at $\Lambda_{exp}$ and $\Lambda_{W}$.

\begin{center}
\renewcommand{\arraystretch}{2}
\begin{table}[!htb]
\hspace{-2.5cm}
\small
\begin{tabular}{|c|c|c|c|c|c|c|c|c|c|c|}
  \hline
  $\epsilon_{P,X}^{l_1 l_2 q_i q_j }$ &$\Lambda_{exp}$&$\Lambda_W$&$S.O, \Lambda_{exp}$&$S.O, \Lambda_{W}$  \\
    \hline
   $\epsilon^{e \mu d s}_{P,X}$ &$2.32\times10^{-7}$&$4.06\times 10^{-7}$&$1.28\times10^{-8}$&$7.82\times10^{-9}$ \\
  \hline
     $\epsilon^{e \mu cu}_{P,X}$ &$1.75\times10^{-3}$&$1.08\times 10^{-3}$&$7.92\times 10^{-5}$&$4.84\times 10^{-5}$\\
  \hline
     $\epsilon^{e \mu bd}_{P,X}$ &$2.35\times10^{-4}$&$1.66\times 10^{-4}$&$5.13\times 10^{-6}$&$3.61\times 10^{-6}$\\
  \hline
       $\epsilon^{e \mu bs}_{P,X}$ &$1.75\times10^{-4}$&$1.23\times 10^{-4}$&$8.27\times 10^{-6}$&$5.83\times 10^{-6}$\\
       \hline
\end{tabular} 
\begin{tabular}{|c|c|c|c|c|c|c|c|c|c|c|}
  \hline
  $\epsilon_{S,X}^{l_1 l_2 q_i q_j }$ &$\Lambda_{exp}$&$\Lambda_W$&$S.O, \Lambda_{exp}$&$S.O, \Lambda_{W}$  \\
    \hline
   $\epsilon^{e \mu d s}_{S,X}$ &$1.05\times10^{-6}$&$5.68\times10^{-7}$&$7.67\times 10^{-7}$&$4.68\times10^{-7}$  \\
  \hline
     $\epsilon^{e \mu cu}_{S,X}$ &$1.34\times10^{-3}$&$8.25\times 10^{-4}$&$1.33\times10^{-3}$&$8.1\times 10^{-4}$ \\
  \hline
     $\epsilon^{e \mu bd}_{S,X}$ &$1.44\times10^{-5}$&$1.01\times 10^{-5}$&$1.44\times 10^{-5}$&$1.01\times 10^{-5}$\\
  \hline
       $\epsilon^{e \mu bs}_{S,X}$ &$2.25\times10^{-5}$&$1.59\times 10^{-5}$&$2.24\times 10^{-5}$&$1.58\times 10^{-5}$\\
  \hline
\end{tabular}

   \caption{Constraints on the dimensionless four-fermion coefficients $\epsilon^{l_1 l_2 q_i q_j }_{P,X}$ and $\epsilon^{l_1 l_2 q_i q_j }_{S,X}$ at the experimental ($\Lambda_{exp}$ for K and D mesons decay and $\Lambda_{m_b}$ for B meson decays) and weak ($\Lambda_{W}$) scale after the RGEs evolution. The last two columns are the sensitivities, or Single Operator (SO) at a time  bounds, see subsection \ref{subsec:SOD}. All bounds apply under permutation of the lepton and/or quark indices.}
\label{table:Scalar}
\end{table}
\renewcommand{\arraystretch}{1}
\end{center}

\renewcommand{\arraystretch}{2}
\begin{table}[!htb]
\small
\hspace{-2.5cm}
\begin{tabular}{|c|c|c|c|c|c|c|c|c|c|c|}
  \hline
  $\epsilon_{A,X}^{l_1 l_2 q_i q_j }$ &$\Lambda_{exp}$&$\Lambda_W$&$S.O, \Lambda_{exp}$&$S.O, \Lambda_{W}$    \\
    \hline
   $\epsilon^{e \mu d s}_{A,X}$ &$5.45\times10^{-6}$&$5.45\times 10^{-6}$&$3.01\times 10^{-7}$&$3.01\times10^{-7}$ \\
  \hline
     $\epsilon^{e \mu cu}_{A,X}$ &$4.51\times10^{-2}$&$4.52\times 10^{-2}$&$2.04\times10^{-3}$&$2.04\times10^{-3}$\\
  \hline
     $\epsilon^{e \mu bd}_{A,X}$ &$1.48\times10^{-2}$&$1.48\times 10^{-2}$&$3.23\times 10^{-4}$&$3.23\times 10^{-4}$\\
  \hline
       $\epsilon^{e \mu bs}_{A,X}$ &$1.11\times10^{-2}$&$1.11\times10^{-2}$&$5.27\times 10^{-4}$&$5.27\times 10^{-4}$\\
  \hline
\end{tabular}
\begin{tabular}{|c|c|c|c|c|c|c|c|c|c|c|c|}
  \hline
  $\epsilon_{V,X}^{l_1 l_2 q_i q_j }$ &$\Lambda_{exp}$&$\Lambda_W$&$S.O, \Lambda_{exp}$&$S.O, \Lambda_{W}$   \\
    \hline
   $\epsilon^{e \mu d s}_{V,X}$ &$4.94\times10^{-6}$&$4.94\times 10^{-6}$&$2.93\times 10^{-6}$&$2.93\times10^{-6}$\\
  \hline
     $\epsilon^{e \mu cu}_{V,X}$ &$1.45\times10^{-3}$&$1.64\times10^{-3}$&$1.39\times10^{-3}$&$1.39\times10^{-3}$\\
  \hline
     $\epsilon^{e \mu bd}_{V,X}$ &$1.49\times10^{-5}$&$1.03\times10^{-4}$&$1.48\times 10^{-5}$&$1.48\times 10^{-5}$\\
  \hline
       $\epsilon^{e \mu bs}_{V,X}$ &$2.56\times10^{-5}$&$8.05\times 10^{-5}$&$2.54\times 10^{-5}$&$2.54\times 10^{-5}$\\
  \hline
\end{tabular}

\vspace{0.5cm}
   \caption{Constraints on the dimensionless four-fermion coefficients $\epsilon^{l_1 l_2 q_i q_j }_{A,X}$ and $\epsilon^{l_1 l_2 q_i q_j }_{V,X}$ at the experimental ($\Lambda_{exp}$ for K and D mesons decay and $\Lambda_{m_b}$ for B meson decays) and weak ($\Lambda_{W}$) scale after the RGEs evolution. The last two columns are the sensitivities, or Single Operator (SO) at a time  bounds, see subsection \ref{subsec:SOD}. All bounds apply under permutation of the lepton and/or quark indices.}
   \label{table:V}
\end{table}
\renewcommand{\arraystretch}{1}

\small
\renewcommand{\arraystretch}{2}
\begin{table}[!htb]
\hspace{2.5cm}
\small
\begin{tabular}{|c|c|c|c|c|c|c|c|c|c|c|}
  \hline
  $\epsilon_{T_X}^{l_1 l_2 q_i q_j }$ &$\Lambda_{exp}$&$\Lambda_W$&$S.O, \Lambda_{exp}$&$S.O, \Lambda_{W}$   \\
    \hline
   $\epsilon^{e \mu d s}_{T_X}$ &$1.23\times10^{-5}$&$1.45\times 10^{-5}$&$8.76\times 10^{-6}$&$1.03\times10^{-5}$\\
  \hline
     $\epsilon^{e \mu cu}_{T_X}$ &$2.01\times10^{-3}$&$2.37\times 10^{-3}$&$1.93\times10^{-3}$&$2.28\times10^{-3}$\\
  \hline
     $\epsilon^{e \mu bd}_{T_X}$ &$2.01\times10^{-5}$&$2.26\times 10^{-5}$&$2\times 10^{-5}$&$2.25\times 10^{-5}$\\
  \hline
       $\epsilon^{e \mu bs}_{T_X}$ &$3.89\times10^{-5}$&$4.37\times 10^{-5}$&$3.87\times 10^{-5}$&$4.35\times 10^{-5}$\\
  \hline
\end{tabular}
   \caption{Constraints on the dimensionless four-fermion coefficients $\epsilon^{l_1 l_2 q_i q_j }_{T_{X}}$  at the experimental ($\Lambda_{exp}$ for K and D mesons decay and $\Lambda_{m_b}$ for B meson decays) and weak ($\Lambda_{W}$) scale after the RGEs evolution. The last two columns are the sensitivities, or Single Operator (SO) at a time  bounds, see subsection \ref{subsec:SOD}. All bounds apply under permutation of the lepton and/or quark indices.}
\label{table:tensor}
\end{table}
\renewcommand{\arraystretch}{1}

\normalsize

\noindent In the leptonic decays, the evolution of the bounds on the pseudoscalar  coefficients between $\Lambda_{exp}$ and  $\Lambda_{W}$ is the most important effect of the RGEs as shown in  the first two columns of the left panel of table \ref{table:Scalar}. As can be seen in eqn. \eqref{REGKaon}, \eqref{REGDmeson} or \eqref{REGBmeson}, the running of the (pseudo)scalar coefficients is $\sim 1.6 (1.4)$, which means that if we  neglect the mixing of the tensor into (pseudo)scalar coefficients, the bounds on $\epsilon_S$ and  $\epsilon_P$ will be better at $\Lambda_W$ for all the decays we considered.  However, the large mixing of the tensor coefficients into the (pseudo)scalar ones  (see eqn. \eqref{ANomdimTS}, \eqref{ANomdimTP} and eqn. \eqref{REGKaon} to \eqref{REGBmeson}) weaken the bounds on pseudoscalar  coefficients at $\Lambda_{W}$  for the Kaon   decay. 
This is due to the fact that the bounds on  $\epsilon_T^{e \mu ds}$  (see the first two columns of table \ref{table:tensor}) are much weaker than the bounds on  $\epsilon_P^{e \mu ds}$  at $\Lambda_{exp}$  (see the   first two columns of the left panel of table \ref{table:Scalar}). Thus, the mixing of $\epsilon_T$ into $\epsilon_P$ will leads to weaker bounds on $\epsilon_P$ at $\Lambda_W$ for the Kaon decay.

\noindent For the D,B and $B_s$ meson decays, the bounds on $\epsilon_P$ are a bit closer to the bound on $\epsilon_T$ at $\Lambda_{exp}$. Even with the large mixing of the tensor into the pseudoscalar coefficients, the bounds on $\epsilon_P^{e \mu cu}$,  $\epsilon_P^{e \mu bd}$ and $\epsilon_P^{e \mu bs}$ will be slightly better at $\Lambda_{W}$ because the running will be larger than the mixing.

\noindent In the semileptonic decays, there is also a mixing between scalar and tensor coefficients, but the bounds on scalar coefficients  at $\Lambda_{W}$ increases a bit  because, similarly to $\epsilon_P^{e \mu cu}$, $\epsilon_P^{e \mu bd}$ and $\epsilon_P^{e \mu bs}$, the bounds on all the scalar coefficients (first two columns of the right panel of table \ref{table:Scalar}) are close to the bounds on the tensor coefficients at $\Lambda_{exp}$. The running of the scalars will be stronger than the mixing of the tensors into the scalars, thus, the bounds on $\epsilon_S$ are better at $\Lambda_{W}$ for all the decays.\\ \\
 For the axial and vector coefficients, there is no running and the mixing is small. The bounds on $\epsilon_A^{e \mu ds}$ and $\epsilon_V^{e \mu ds}$ at $\Lambda_{exp}$ are very close (see table \ref{table:V}), this explains why there is no evolution of these bounds at $\Lambda_{W}$.   However, for the D, B and $B_s$ decays, the bounds on $\epsilon_A$ are much weaker than the bounds on $\epsilon_V$ at  $\Lambda_{exp}$, especially for the B and  $B_s$ decay. Thus, the bounds on $\epsilon_A^{e \mu cu}$, $\epsilon_A^{e \mu bd}$ and $\epsilon_A^{e \mu bs}$ do not evolve significantly at $\Lambda_{W}$, but the mixing of the axial into vector coefficients will lead to weaker bounds on $\epsilon_V^{e \mu cu}$, $\epsilon_V^{e \mu bd}$ and $\epsilon_V^{e \mu bs}$ at $\Lambda_{W}$ as shown in the first two columns of the two panels  of   table \ref{table:V}. \\
 Finally, the running of tensor coefficients is tiny, and the mixing of the (pseudo)scalar coefficients into the tensor ones is small. Thus, the evolution of the bounds is  small for the tensor coefficients (first two columns of table \ref{table:tensor}) as for  the bounds on vector and axial coefficients in the Kaon decay (first two columns of table \ref{table:V}). Finally, the matching at $\Lambda_{W}$ along with the evolution of the bounds between $\Lambda_{W}$ and $\Lambda_{NP}$ will be given in a future publication \cite{futureASSD}. \\
 
 \subsection{Single operator approximation}
 \label{subsec:SOD}

We also computed the sensitivities of the
   various decays to the coefficients  at $\Lambda_{exp}$,
   and these  are given in the third column of tables \ref{table:Scalar}
   to \ref{table:tensor}.
   The sensitivity is the value of the coefficient below which
   it could not  have been observed, and is  calculated
   as a ``Single Operator'' (SO) at a time bound, 
   that is by allowing only one non-zero coefficient
   at a time in the branching ratio (see eqn \eqref{leptonic rate} and \eqref{M3body}). This is different from setting bounds on coefficients (first two columns of table \ref{table:Scalar} to \ref{table:tensor}), which are obtained with all coefficients non-zero, and  exclude the parameter space outside the allowed range.  It is clear that the sensitivities are sometimes an excellent
   approximation to the bounds, and sometimes differ by
 orders of magnitude. \\
To compute the evolution of the sensitivities of the decays to the coefficients at $\Lambda_{W}$ (given in the last column of table  \ref{table:Scalar} to \ref{table:tensor}), we still kept only one non-zero coefficients at $\Lambda_{exp}$ and considered  only the running of the coefficients (the diagonal terms in eqn. \eqref{REGKaon} to \eqref{REGBmeson}). For example, computing the sensitivity of the leptonic Kaon decay to a pseudoscalar coefficient  at $\Lambda_{W}$ in the SO approximation requires to multiply the first term in eqn.  \eqref{covmatelemt2} by the first (or third)  diagonal term squared  in eqn. \eqref{REGKaon}. Then, inverting the product and taking the square root  will give the sensitivity of the decay to the coefficient at  $\Lambda_{W}$.

 \subsection{Updating the bounds}
 \label{subsec:update}
 
In future years, the experimental data
   on LFV meson decays   could improve, so  in this section,
   we consider how  to update our bounds, 
   without inverting large matrices.\\
\\
 The bounds on  coefficients at $\Lambda_{exp}$ 
obtained in this work 
 are of the form $|\epsilon| <\sqrt{ BR^{exp}}\times$ constant. Thus,
 all the bounds at $\Lambda_{exp}$ given in tables \ref{table:Scalar} to \ref{table:tensor} can be  updated by  rescaling by $\sqrt{ (BR_{new}^{exp})/(BR_{old}^{exp})}$ when the data   improves. However, in principle,
 the 10$\times$10 matrix of eqn (\ref{M-1prime})  must then be inverted
 to obtain the bounds  at $\Lambda_{W}$.  So we now
 describe approximations that allow to obtain  the
  bounds  at $\Lambda_{W}$  with   manageable matrices.\\ 
  \\ 
  For the  semileptonic decay, the bounds at
   $\Lambda_{exp}$ can be obtained 
   by neglecting  all the interference terms 
   between the  scalar, vector and tensor coefficients of
   either  chirality 
   (see eqn. \eqref{M3body}). 
   The $6 \times 6$ matrix in eqn. \eqref{M66} then becomes diagonal
   and easy to invert.
   This approximation will give  bounds at $\Lambda_{exp}$ on
   $\epsilon_{S,X}$,  $\epsilon_{V,X}$ and  $\epsilon_{T,X}$
   close to those obtained in the first  column  of tables
   \ref{table:Scalar} to \ref{table:tensor} (which
   include  the interference terms).
  \\ 
\\
    In the leptonic decay  (eqn. \eqref{leptonic rate}),
 a reasonable  approximation for the bounds at $\Lambda_{exp}$
 is to keep the interference between axial and
    pseudoscalar coefficients  of opposite chirality
    (with $m_2 = m_{\mu}$ in eqn. \eqref{leptonic rate}).
    The other interference terms, proportional to  $m_1 = m_e $,
    can be neglected. 
    Thus, bounds on  $\epsilon_{A}$ and  $\epsilon_{P}$ at $\Lambda_{exp}$,
    which are a reasonable approximation to the first
    column of tables \ref{table:Scalar} and \ref{table:V}, can be obtained by
 inverting a  $2\times2$ matrix in the basis $\left( \epsilon_{P,X},\epsilon_{A,Y}\right)$ where $X \in L,R$ and $Y \in R,L$, instead of the $4\times4$ matrix in eqn. \eqref{M44}.   \\
\\
\\
To obtain bounds at $\Lambda_{W}$, it is necessary to keep the mixing between $\epsilon_{S}$,  $\epsilon_{P}$, $\epsilon_{T}$,
and between $\epsilon_{V}$ and $\epsilon_{A}$.  Then, the bounds on $\epsilon_{S}$, $\epsilon_{P}$, $\epsilon_{T}$, $\epsilon_{V}$ and $\epsilon_{A}$   at $\Lambda_{W}$ can be obtained by considering $M^{-1'}$ in eqn. \eqref{M-1prime}  as a product  of $5\times5$ matrices in the basis ($\epsilon_{P,X}$, $\epsilon_{S,X}$, $\epsilon_{T,X}$,$\epsilon_{V,Y}$,$\epsilon_{A,Y}$) where X and Y are the chirality. However,  $\epsilon_{S}$,  $\epsilon_{P}$ and $\epsilon_{T}$  must have the same chirality, but different from the chirality of $\epsilon_{V}$ and $\epsilon_{A}$ in order to take into account the mixing induced by the RGEs, that occurs only for coefficients of the same chirality (see eqn. \eqref{RGE}, and \eqref{REGKaon} to \eqref{REGBmeson}). This is due to the fact that it is necessary to keep the interference between axial and pseudoscalar coefficients of different chiralities to compute the bounds on  $\epsilon_{P,X}$ and $\epsilon_{A,Y}$.

\section{Conclusion}

\noindent In this paper,
we consider  operators which simultaneously change lepton and quark flavour,
and obtain constraints on the coefficients 
using available data on (semi)leptonic pseudoscalar meson decays.
Section \ref{sec:2}  lists the dimension six, two lepton two quark operators
and their associated coefficients at   the experimental scale $\Lambda_{exp}$.
Scalar, pseudoscalar, vector, axial and tensor operators are included.
The leptonic and semileptonic branching ratios of pseudoscalar mesons, as a function of the operator coefficients, are given in section \ref{sec:3}.
We find tensor operators do not contribute to the leptonic decays but only to the semileptonic decays, in which the interference between $\epsilon_{S,L}$ ($\epsilon_{S,R}$) and $\epsilon_{T_R}$ ($\epsilon_{T_L}$) vanishes.
The constraints on operator coefficients, evaluated at the
experimental scale, are given in
tables \ref{table:Scalar}, \ref{table:V} and \ref{table:tensor}
and discussed in  section \ref{sec:4}. The bounds are obtained
via the appropriate covariance matrices,  which allows to take into account the interferences among operators (see eqn. \eqref{leptonic rate},\eqref{M3body},\eqref{M44} and \eqref{M66}). The  matrices are given  in appendix B.
Section \ref{sec:5} gives  the  Renormalization Group  evolution of the coefficients
from the experimental to the weak scale $\Lambda_{W}$, and  the formalism
used to compute the covariances matrices at $\Lambda_{W}$. The
weak-scale constraints on the coefficients are given in tables
 \ref{table:Scalar}, \ref{table:V} and \ref{table:tensor}.
The large mixing of tensor coefficients into  (pseudo)scalar   coefficients
has important consequences on the evolution of the bounds on scalar
and pseudoscalar coefficients.
Indeed,  in the case of the  kaon   decay,
the  experimental-scale bounds on tensor coefficients
are significantly weaker than
those  on pseudoscalars. As a result,
the  pseudoscalar bounds are weaker  at $\Lambda_{W}$,
compared to the bounds at $\Lambda_{exp}$.  The bounds on scalar coefficients at $\Lambda_{W}$ are slightly stronger than at  $\Lambda_{exp}$.  There is no running for the vector and axial coefficients, due to the fact we consider quark-flavor changing operators,
and the mixing is small, but the bounds on axial coefficients are much weaker than the bounds on vector coefficients for the D, B and $B_s$ decays, this leads to much weaker bounds on vector coefficients at $\Lambda_{W}$.  Similarly,
the running and mixing of tensor coefficients are small.
As a result,  the bounds  on the  axial and tensor  coefficients
do not evolve significantly between
the experimental and weak scales.

We conclude by noting the importance of
including interferences among operators in computing  the bounds on
their coefficients. As shown in subsection \ref{subsec:SOD}, the sensitivities of the decays to $\epsilon_{P}$ and $\epsilon_{A}$  obtained at $\Lambda_{exp}$ and to $\epsilon_{P}$, $\epsilon_{A}$ and $\epsilon_{V}$ at   $\Lambda_{W}$ in the single operator approximation are better by several orders of magnitude compared to the bounds obtained by keeping the interferences among operators. We found that the Renormalization
Group running between the experimental and weak scales
has an important effect on the evolution of the bounds,
especially the large mixing of the tensor (axial) into the pseudoscalar (vector),
which lead to  weaker bounds on pseudoscalar (vector) coefficients
at $\Lambda_{W}$ for the Kaon (D, B and $B_s$)  decay.     

\newpage

\begin{appendices}

\section{Constants}
\label{appendix:Constants}
 \noindent In this section, we give all the constants used in our calculations : \\

\hspace{-2cm}
\small
 \begin{tabular}{|c|c|c|c|c|c|c|c|c|}
  \hline
  $P_1$&$K^0_L$ & $K^+$& $D^0$ &$D^+$ & $D^+_S$ & $B^0$&$B^0_S$&$B^+$ \\
  \hline
  $f_{P_1} (MeV)$&155.72 \cite{Kcstdecay,KDform} & 155.6 \cite{Kcstdecay,KDform}& 211.5 \cite{Kcstdecay,BandDdecaycst} & 212.6 \cite{Kcstdecay,BandDdecaycst}& 249.8 \cite{BandDdecaycst} &190.9 \cite{Kcstdecay}&230.7\cite{BandDdecaycst}   &187.1 \cite{Kcstdecay}\\
  \hline
  $f_+^{P_1 \pi}(0)$&0.966 \cite{KDform}  & 0.966 \cite{KDform}  & 0.666 \cite{KDform} & 0.666 \cite{KDform}&0.666 \cite{KDform} &0.25 \cite{Bform}  &0.25 \cite{Bform} &0.25 \cite{Bform} \\ 
  \hline
   $f_+^{P_1 K}(0)$&- & - & 0.747 \cite{KDform}& 0.747 \cite{KDform}&0.747 \cite{KDform} &0.31 \cite{Bform} &0.31 \cite{Bform}&0.31 \cite{Bform}  \\ 
  \hline
   $\lambda_+$& $2.82 \times 10^{-2}$ \cite{PDG} &$2.97 \times 10^{-2}$ \cite{PDG}  & -&-&-&-&-&- \\ 
  \hline
   $\lambda_0$& $1.8 \times 10^{-2}$ \cite{PDG}& $1.95 \times 10^{-2}$ \cite{PDG}   & -&-&-&- &-&- \\ 
  \hline
\end{tabular}\\

\normalsize

\noindent All the masses and lifetime can be found in \cite{PDG}.
 
\section{Kinematics and form factors for semileptonic decays }
\label{appendix:kin-formfactor} 

\noindent In this section, we give the form factor and the detailed scalar product of eqn. \eqref{M3body}. \\

\noindent The  $q^2$ dependence of the form factors for the Kaon is given by \cite{Khadroelemevol} : 

\begin{equation}
f_{+,0}^{K \pi}(q^2) = f_+^{K \pi}(0) \left( 1+ \lambda_{+,0}\frac{q^2}{M_{\pi}^2} \right);\quad f_-^{K \pi}(q^2) = f_+^{K \pi}(0)(\lambda_{0}-\lambda_{+})\frac{M_{K^+}^2-M_{\pi^+}^2}{M_{\pi^+}^2}
\end{equation}

\noindent and for the D and B mesons, are given by \cite{Dhadroelemevol,Bhadroelemevol} : 
\begin{equation}
f_+(q^2) =  \frac{f_+(0)}{1-q^2/m_{1^-}^2} ;\quad f_0(q^2) = \frac{f_0(0)}{1-q^2/m_{0^+}^2}; \quad 
f_-(q^2)=(f_0(q^2)-f_+(q^2))\frac{M^2-m_3^2}{q^2}
\end{equation}

\noindent where $\lambda_{+,0}$ are constants, $m_{J^P}$  is the mass of the lightest resonance with the right quantum numbers to mediate the transition ($D_s^+$ and $D_s^{*+}$ for example). We took $q^2  = q^2_{\text{max}} =(M - m_3)^2$ to compute the form factors $f_+$, $f_-$ and $f_0$. All these values can be found in appendix \ref{appendix:Constants}.


\noindent Finally, the scalar product in eqn. \eqref{M3body} can be written as functions of the two kinematical variables $q^2$ and $\cos \theta$ \cite{PDG,Ilisie} in the phase space integrals of eqn. \eqref{3 body rate phase space}. \\

\begin{equation}
p_1.p_2 = \frac{q^2  - m_1^2 -m_2^2}{2}, \quad p_1.q = \frac{q^2 + m_1^2 - m_2^2}{2}, \quad p_2.q = \frac{q^2  + m_2^2 - m_1^2 }{2}
\end{equation}

\begin{equation}
p_3.q = \frac{M^2 - m_3^2 -q^2}{2}, \quad  p_1.p_3 = p_3.q -p_2.p_3, \quad p_1.P =  p_1.q + 2p_1.p_3, \quad p_2.P =  p_2.q + 2p_2.p_3
\end{equation}

\begin{equation}
p_2.p_3 =  \frac{1}{4q^2}
( M^2 -m_3^2-q^2  )(q^2  + m_2^2 -m_1^2) + \frac{1}{4q^2}\sqrt{\lambda(M^2,m_3^2,q^2)}\sqrt{\lambda(q^2,m_1^2,m_2^2)}\cos\theta 
\end{equation}

\begin{equation}
 k.p_3 = \frac{M^2 + m_3^2 - q^2}{2},\quad P.q = M^2 - m_3^2, \quad P^2 = 2M^2 + 2m_3 - q^2
\end{equation} \\

\section{RGEs}
\label{appendix:RGESmatrices}

\noindent In this section, we give the $10\times10$ matrices obtained with eqn. \eqref{RGE} we used to obtained the bounds at $\Lambda_W$ (with eqn. \eqref{M-1prime}).\\
\noindent For the decay of light quark (Kaon and D meson decays), the experimental scale is taken as 2 GeV because most of the time, it's the renormalization scale chosen to obtain the lattice form factors.\\

\noindent The evolution of the coefficients ($\epsilon^{e \mu ds}$) involved in the Kaon decays is given by :

\footnotesize
\begin{equation}
\hspace{-1.8cm}
\left(\begin{array}{c}\epsilon_{P,L}\\\epsilon_{A,L}\\\epsilon_{P,R}\\\epsilon_{A,R}\\\epsilon_{S,L}\\\epsilon_{V,L}\\\epsilon_{T_{L}} \\ \epsilon_{S,R} \\\epsilon_{V,R}\\\epsilon_{T_{R}}\\\end{array} \right)_{\Lambda_{exp}} = 
\left(
\begin{array}{cccccccccc}
 1.64 & 0 & 0 & 0 & 0 & 0 & -0.0429 & 0 & 0 & 0 \\
 0 & 1 & 0 & 0 & 0 & 0.00857 & 0 & 0 & 0 & 0 \\
 0 & 0 & 1.64 & 0 & 0 & 0 & 0 & 0 & 0 & 0.0429 \\
 0 & 0 & 0 & 1 & 0 & 0 & 0 & 0 & -0.00857 & 0 \\
 0 & 0 & 0 & 0 & 1.64 & 0 & 0.0429 & 0 & 0 & 0 \\
 0 & 0.00857 & 0 & 0 & 0 & 1 & 0 & 0 & 0 & 0 \\
 -0.00162 & 0 & 0 & 0 & 0.00162 & 0 & 0.849 & 0 & 0 & 0 \\
 0 & 0 & 0 & 0 & 0 & 0 & 0 & 1.64 & 0 & 0.0429 \\
 0 & 0 & 0 & -0.00857 & 0 & 0 & 0 & 0 & 1 & 0 \\
 0 & 0 & 0.00162 & 0 & 0 & 0 & 0 & 0.00162 & 0 & 0.849 \\
\end{array}
\right)
\left(\begin{array}{c}\epsilon_{P,L}\\\epsilon_{A,L}\\\epsilon_{P,R}\\\epsilon_{A,R}\\\epsilon_{S,L}\\\epsilon_{V,L}\\\epsilon_{T_{L}} \\ \epsilon_{S,R} \\\epsilon_{V,R}\\\epsilon_{T_{R}}\\\end{array} \right)_{\Lambda_{W}}
\label{REGKaon}
\end{equation} \\

\normalsize
\noindent For the D meson decays, the evolution of the coefficients ($\epsilon^{e \mu cu}$) is given by :

\footnotesize
\begin{equation}
\hspace{-1.8cm}
\left(\begin{array}{c}\epsilon_{P,L}\\\epsilon_{A,L}\\\epsilon_{P,R}\\\epsilon_{A,R}\\\epsilon_{S,L}\\\epsilon_{V,L}\\\epsilon_{T_{L}} \\ \epsilon_{S,R} \\\epsilon_{V,R}\\\epsilon_{T_{R}}\\\end{array} \right)_{\Lambda_{exp}} = 
\left(
\begin{array}{cccccccccc}
 1.64 & 0 & 0 & 0 & 0 & 0 & 0.0857 & 0 & 0 & 0 \\
 0 & 1 & 0 & 0 & 0 & -0.0171 & 0 & 0 & 0 & 0 \\
 0 & 0 & 1.64 & 0 & 0 & 0 & 0 & 0 & 0 & -0.0857 \\
 0 & 0 & 0 & 1 & 0 & 0 & 0 & 0 & 0.0171 & 0 \\
 0 & 0 & 0 & 0 & 1.64 & 0 & -0.0857 & 0 & 0 & 0 \\
 0 & -0.0171 & 0 & 0 & 0 & 1 & 0 & 0 & 0 & 0 \\
 0.00325 & 0 & 0 & 0 & -0.00325 & 0 & 0.847 & 0 & 0 & 0 \\
 0 & 0 & 0 & 0 & 0 & 0 & 0 & 1.64 & 0 & -0.0857 \\
 0 & 0 & 0 & 0.0171 & 0 & 0 & 0 & 0 & 1 & 0 \\
 0 & 0 & -0.00325 & 0 & 0 & 0 & 0 & -0.00325 & 0 & 0.847 \\
\end{array}
\right)
\left(\begin{array}{c}\epsilon_{P,L}\\\epsilon_{A,L}\\\epsilon_{P,R}\\\epsilon_{A,R}\\\epsilon_{S,L}\\\epsilon_{V,L}\\\epsilon_{T_{L}} \\ \epsilon_{S,R} \\\epsilon_{V,R}\\\epsilon_{T_{R}}\\\end{array} \right)_{\Lambda_{W}}
\label{REGDmeson}
\end{equation} \\

\normalsize
\noindent \noindent In the B and $B_s$  meson decay, the reference scale is the b quark mass  ($\Lambda_{m_b}\sim 4.18$ GeV). Thus, the evolution of the coefficients ($\epsilon^{e \mu bd}$ and $\epsilon^{e \mu bs}$) is slightly smaller. \\
In fact, in eqn. \eqref{RGE}, the part with the anomalous dimension  that gives the matrix element in eqn. \eqref{REGKaon} is multiplied by a factor $\log(\frac{\Lambda_W}{\Lambda_{m_b}})/\log(\frac{\Lambda_W}{\Lambda_{exp}}) \sim 0.8$. Moreover, the strong coupling constant at $\Lambda_{m_b}$  will also be smaller ($\alpha_s(\Lambda_{m_b}) \sim 0.23$ and $\alpha_s(\Lambda_{exp}) \sim 0.3$). Thus, for the B and $B_s$ meson decays, the evolution of the coefficients ($\epsilon^{e \mu bd}$ and $\epsilon^{e \mu bs}$ ) is given by :

\footnotesize
\begin{equation}
\hspace{-1.8cm}
\left(\begin{array}{c}\epsilon_{P,L}\\\epsilon_{A,L}\\\epsilon_{P,R}\\\epsilon_{A,R}\\\epsilon_{S,L}\\\epsilon_{V,L}\\\epsilon_{T_{L}} \\ \epsilon_{S,R} \\\epsilon_{V,R}\\\epsilon_{T_{R}}\\\end{array} \right)_{\Lambda_{exp}} = 
\left(
\begin{array}{cccccccccc}
 1.42 & 0 & 0 & 0 & 0 & 0 & -0.0317 & 0 & 0 & 0 \\
 0 & 1 & 0 & 0 & 0 & 0.00686 & 0 & 0 & 0 & 0 \\
 0 & 0 & 1.42 & 0 & 0 & 0 & 0 & 0 & 0 & 0.0317 \\
 0 & 0 & 0 & 1 & 0 & 0 & 0 & 0 & -0.00686 & 0 \\
 0 & 0 & 0 & 0 & 1.42 & 0 & 0.0317 & 0 & 0 & 0 \\
 0 & 0.00686 & 0 & 0 & 0 & 1 & 0 & 0 & 0 & 0 \\
 -0.00126 & 0 & 0 & 0 & 0.00126 & 0 & 0.890 & 0 & 0 & 0 \\
 0 & 0 & 0 & 0 & 0 & 0 & 0 & 1.42 & 0 & 0.0317 \\
 0 & 0 & 0 & -0.00686 & 0 & 0 & 0 & 0 & 1 & 0 \\
 0 & 0 & 0.00126 & 0 & 0 & 0 & 0 & 0.00126 & 0 & 0.890 \\
\end{array}
\right)
\left(\begin{array}{c}\epsilon_{P,L}\\\epsilon_{A,L}\\\epsilon_{P,R}\\\epsilon_{A,R}\\\epsilon_{S,L}\\\epsilon_{V,L}\\\epsilon_{T_{L}} \\ \epsilon_{S,R} \\\epsilon_{V,R}\\\epsilon_{T_{R}}\\\end{array} \right)_{\Lambda_{W}}
\label{REGBmeson}
\end{equation} \\

\section{Covariances matrix}
\label{appendix:M2/M3-1}

 \noindent In this section, we give details of the formalism introduced in section \ref{sec:4}, eqn. \eqref{CMmpoins1C}. The matrices in the basis    $\left( \epsilon_{P,L},\epsilon_{A,L}, \epsilon_{P,R} ,\epsilon_{A,R}\right)$ and  $\left(\epsilon_{S,L} ,\epsilon_{V,L},\ \epsilon_{T_L},\ \epsilon_{S,R} ,\epsilon_{V,R},\ \epsilon_{T_R}\right)$  are written : \\

\begin{align}
&M_2^{-1}=\frac{1}{BR_2^{exp}}
\left[
\begin{array}{cccccccccccc} 
& SP_+' & \frac{1}{2} SP_+VA_+'&\frac{1}{2}SP_+SP_-'  &\frac{1}{2}SP_+VA_-'     \\
&\frac{1}{2} SP_+VA_+'  & VA_-'&\frac{1}{2}SP_-VA_+'  &\frac{1}{2}VA_+VA_-'    \\
&\frac{1}{2}SP_+SP_-' &\frac{1}{2}SP_-VA_+' &SP_-' &\frac{1}{2}SP_-VA_-' \\
&\frac{1}{2}SP_+VA_-' &\frac{1}{2}VA_+VA_-'    &\frac{1}{2}SP_-VA_-' &VA_+' \\
\end{array}
\right]
\label{M44}
\end{align} \\

\begin{align}
&M_3^{-1}=\frac{1}{BR_3^{exp}}
\left[
\begin{array}{cccccccccccc} 
& SP_+ & \frac{1}{2} SP_+VA_- &\frac{1}{2}SP_+T_+ &\frac{1}{2}SP_+SP_-  &\frac{1}{2}SP_+VA_+& \frac{1}{2}SP_+T_-    \\
&\frac{1}{2} SP_+VA_-  & VA_-& \frac{1}{2}VA_-T_+  &\frac{1}{2}SP_-VA_-  &\frac{1}{2}VA_+VA_- &\frac{1}{2}VA_-T_-   \\
&\frac{1}{2}SP_+T_+ & \frac{1}{2}VA_-T_+  &T_+ &\frac{1}{2}SP_-T_+ &\frac{1}{2}VA_+T_+ &\frac{1}{2}T_+T_- \\
&\frac{1}{2}SP_+SP_- &\frac{1}{2}SP_-VA_-    &\frac{1}{2}SP_-T_+ & SP_- & \frac{1}{2} SP_-VA_+ &\frac{1}{2}SP_-T_-  \\
&\frac{1}{2}SP_+VA_+ &\frac{1}{2}VA_+VA_-  &\frac{1}{2}VA_+T_+ & \frac{1}{2} SP_-VA_+ &VA_+ & \frac{1}{2}VA_+T_- \\
&\frac{1}{2}SP_+T_- &\frac{1}{2}VA_-T_- &\frac{1}{2}T_+T_- &\frac{1}{2}SP_-T_-  &\frac{1}{2}VA_+T_-  &T_- \\
\end{array}
\right]
\label{M66}
\end{align} \\

\noindent

\noindent Inverting $M_2^{-1}$ [$M_3^{-1}$]  will give the bounds on the coefficients involved in the leptonic [semileptonic] decays. Finally, note that for semileptonic Kaon and  D meson decays, the experimental upper limit are not the same for $\mu^+ e^-$ and $\mu^- e^+$ in the final state. In this case, we sum the  $M_3^{-1}$ for each bound and then invert it to obtain the covariance matrix of section \ref{sec:4}. The matrix elements of eqn. \eqref{M44} are written : 

\begin{equation}
\begin{split}
SP_+' =SP_-' &=  C_{2\text{body}} \tilde{P'}^2 (P_1^2 -m_i^2-m_j^2)\\
VA_-' =VA_+'&= C_{2\text{body}} \tilde{A'}^2[(P_1^2-m_i^2-m_j^2) (m_i^2+m_j^2) +4m_i^2m_j^2] \\
SP_+VA_-' &=SP_-VA_+'= -2C_{2\text{body}}\tilde{P'}\tilde{A'}
m_j(P_1^2 +m_i^2 -m_j^2) \\
SP_+VA_+' &=SP_-VA_-'= 2C_{2\text{body}}\tilde{P'}\tilde{A'}
m_i(P_1^2 +m_j^2 -m_i^2) \\
SP_+SP_-' &= -4C_{2\text{body}}\tilde{P'}^2m_jm_i \\
VA_+VA_-' &= -4C_{2\text{body}}\tilde{A'}^2P_1^2m_jm_i \\
C_{2\text{body}} &=\frac{\tau_{P_1}r^*G_F^2}{\pi P_1^2}
\end{split}
\label{covmatelemt2}
\end{equation}

\noindent For simplicity we note $d \phi = \int_{(m_1+m_2)^2}^{(M-m_3)^2}dq^2 \int_{-1}^{1}d\cos\theta \frac{\sqrt{\lambda(M^2,m_3^2,q^2)}\sqrt{\lambda(q^2,m_1^2,m_2^2)}}{q^2} $ and the matrix elements of eqn.  \eqref{M66} are written :

\begin{equation}
\hspace{-2cm}
\begin{split}
SP_+ = SP_- &= 2C_{3\text{body}} \tilde{S}^2 (p_1.p_2)d \phi \\
VA_+ = VA_-&=\frac{1}{4}C_{3\text{body}} [f_+^2 \left(
 4(p_1.P)(p_2.P) - 2P^2 (p_1.p_2) \right) + f_-^2 \left(
 4(p_1.q)(p_2.q) - 2q^2 (p_1.p_2) \right) \\
  &+ 4f_+f_-   \left((p_1.q)(p_2.P) + (p_1.P)(p_2.q)-(p_1.p_2)(P.q)    \right)] d \phi\\
T_+ = T_- &= 4C_{3\text{body}}\tilde{T'}^2 [ 4 (p_1.q)(p_2.P)(P.q) + 4(p_1.P)(p_2.q)(P.q) -2(p_1.p_2)(P.q)^2  \\
 &+2P^2q^2(p_1.p_2) - 4 P^2(p_1.q)(p_2.q) - 4q^2(p_1.P)(p_2.P) 
 ]  d \phi\\
 SP_+VA_- &= SP_-VA_+ = -2C_{3\text{body}} \tilde{S}m_2[ \left(f_+ (p_1.P) + f_-(p_1.q) \right)] d \phi\\
 SP_+VA_+ &= SP_-VA_- = 2C_{3\text{body}}\tilde{S}m_1[ \left(f_+ (p_2.P) + f_-(p_2.q) \right) ] d \phi \\
 SP_+SP_- &= -4 C_{3\text{body}}\tilde{S}^2 m_1 m_2 d \phi \\
 VA_+VA_-&=-C_{3\text{body}} m_1 m_2[f_-^2q^2 + f_+^2P^2 + 2 f_+ f_-(P.q)]d \phi \\ T_+T_-&=16C_{3\text{body}} \tilde{T'}^2 m_1 m_2[(P.q)^2-P^2q^2]d \phi   \\
 SP_+T_+ &= SP_-T_-= 8C_{3\text{body}} \tilde{S}\tilde{T}[\left((p_1.P)(p_2.q)-(p_1.q)(p_2.P)  \right) ]d \phi \\
 SP_+T_- &= SP_-T_+=0 \\
VA_+T_- &= VA_-T_+ =  4C_{3\text{body}}\tilde{T'}m_2[f_+((p_1.q)p^2-(P.p_1)(P.q)) +  f_-((p_1.q)(P.q)-(p_1.P)q^2)]d \phi \\
VA_+T_+ &= VA_-T_- =  4C_{3\text{body}}  \tilde{T'}m_1[ \left(f_+((P^2)(p_2.q)-(p_2.P)(P.q)) +f_-((p_2.q)(P.q) -(q^2)(p_2.P)) \right)]d \phi  \\
{C_{3\text{body}}} &=  \frac{\tau_{P_1}}{\pi^3}\frac{8G_F^2}{512M^3} \\ 
\end{split}
\label{covmatelemt3}
\end{equation}

\section{Covariances matrices at $\Lambda_{exp}$ and $\Lambda_W$}
\label{appendix:bounds exp/w}

\noindent In this section, we give the covariance matrix at $\Lambda_{exp}$ and at $\Lambda_W$, after the RGEs evolution.

\subsection*{Kaon decays}

\noindent Using the upper limit of table \ref{table:exptal bounds}, for the leptonic Kaon decay, we compute the associated covariance matrix in the basis ($\epsilon_{P,L}^{e \mu ds},\epsilon_{A,L}^{e \mu ds},  \epsilon_{P,R}^{e \mu ds} \epsilon_{A,R}^{e \mu ds}$) :  \\

\begin{equation}
\left(
\begin{array}{cccc}
 5.38\times 10^{-14} & -2.33\times 10^{-14} & -1.25\times 10^{-15} & 1.26\times 10^{-12} \\
 -2.33\times 10^{-14} & 2.97\times 10^{-11} & 1.26\times 10^{-12} & -4.03\times 10^{-13} \\
 -1.25\times 10^{-15} & 1.26\times 10^{-12} & 5.38\times 10^{-14} & -2.33\times 10^{-14} \\
 1.26\times 10^{-12} & -4.03\times 10^{-13} & -2.33\times 10^{-14} & 2.97\times 10^{-11} \\
\end{array}
\right)
\end{equation} \\

\noindent Then we use the bounds on semileptonic Kaon decay to compute the covariance matrix for the semileptonic decays in the basis ($\epsilon_{S,L}^{e \mu ds},\epsilon_{V,L}^{e \mu ds} , \epsilon_{T_L}^{e \mu ds}, \epsilon_{S,R}^{e \mu ds},\epsilon_{V,R}^{e \mu ds}, \epsilon_{T_R}^{e \mu ds})$  : \\ \\

\begin{equation}
\hspace{-1cm}
\left(
\begin{array}{cccccc}
 1.09\times 10^{-12} & 3.51\times 10^{-12} & 6.11\times 10^{-12} & 1.39\times 10^{-14} & 1.96\times 10^{-13} & 7.49\times 10^{-13} \\
 3.51\times 10^{-12} & 2.44\times 10^{-11} & 4.26\times 10^{-11} & 1.96\times 10^{-13} & 2.10\times 10^{-12} & 6.50\times 10^{-12} \\
 6.11\times 10^{-12} & 4.26\times 10^{-11} & 1.51\times 10^{-10} & 7.49\times 10^{-13} & 6.50\times 10^{-12} & 1.58\times 10^{-11} \\
 1.39\times 10^{-14} & 1.96\times 10^{-13} & 7.49\times 10^{-13} & 1.09\times 10^{-12} & 3.51\times 10^{-12} & 6.11\times 10^{-12} \\
 1.96\times 10^{-13} & 2.10\times 10^{-12} & 6.50\times 10^{-12} & 3.51\times 10^{-12} & 2.44\times 10^{-11} & 4.26\times 10^{-11} \\
 7.49\times 10^{-13} & 6.50\times 10^{-12} & 1.58\times 10^{-11} & 6.11\times 10^{-12} & 4.26\times 10^{-11} & 1.51\times 10^{-10} \\
\end{array}
\right)
\end{equation} \\

\noindent The diagonal elements give the bounds on $|\epsilon|^2$. The bounds on the coefficients are the square root of the  diagonal elements. For instance,  $\epsilon^{e \mu d s}_{S,L}$ is  excluded above $\sqrt{1.09 \times 10^{-12}}$.

\noindent The covariance matrix in the basis $\left( \begin{array}{c}\epsilon_{P,L}^{e \mu ds},\epsilon_{A,L}^{e \mu ds},\epsilon_{P,R}^{e \mu ds},\epsilon_{A,R}^{e \mu ds},\epsilon_{S,L}^{e \mu ds},\epsilon_{V,L}^{e \mu ds},\epsilon_{T_{L}}^{e \mu ds}, \epsilon_{S,R}^{e \mu ds},\epsilon_{V,R}^{e \mu ds},\epsilon_{T_{R}}^{e \mu ds}\\\end{array} \right)_{\Lambda_W}$ is :

\tiny
\begin{equation}
\hspace{-2.5cm}
\left(
\begin{array}{cccccccccc}
 1.64\times 10^{-13} & -2.55\times 10^{-14} & -1.55\times 10^{-14} & 7.73\times 10^{-13} & -2.91\times 10^{-14} & 1.31\times 10^{-12} & 5.51\times 10^{-12} & -9.15\times 10^{-16} & 2.07\times 10^{-13} &
   5.75\times 10^{-13} \\
 -2.55\times 10^{-14} & 2.97\times 10^{-11} & 7.73\times 10^{-13} & -4.03\times 10^{-13} & -7.10\times 10^{-15} & -4.64\times 10^{-13} & -4.30\times 10^{-13} & 7.35\times 10^{-16} & -2.15\times 10^{-14}
   & -6.72\times 10^{-14} \\
 -1.55\times 10^{-14} & 7.73\times 10^{-13} & 1.64\times 10^{-13} & -2.55\times 10^{-14} & 9.15\times 10^{-16} & -2.07\times 10^{-13} & -5.75\times 10^{-13} & 2.91\times 10^{-14} & -1.31\times 10^{-12} &
   -5.51\times 10^{-12} \\
 7.73\times 10^{-13} & -4.03\times 10^{-13} & -2.55\times 10^{-14} & 2.97\times 10^{-11} & -7.35\times 10^{-16} & 2.15\times 10^{-14} & 6.72\times 10^{-14} & 7.10\times 10^{-15} & 4.64\times 10^{-13} &
   4.30\times 10^{-13} \\
 -2.91\times 10^{-14} & -7.10\times 10^{-15} & 9.15\times 10^{-16} & -7.35\times 10^{-16} & 3.22\times 10^{-13} & 8.29\times 10^{-13} & -1.11\times 10^{-12} & -8.03\times 10^{-15} & -8.12\times 10^{-14}
   & -3.49\times 10^{-14} \\
 1.31\times 10^{-12} & -4.64\times 10^{-13} & -2.07\times 10^{-13} & 2.15\times 10^{-14} & 8.29\times 10^{-13} & 2.44\times 10^{-11} & 5.02\times 10^{-11} & -8.12\times 10^{-14} & 2.10\times 10^{-12} &
   7.66\times 10^{-12} \\
 5.51\times 10^{-12} & -4.30\times 10^{-13} & -5.75\times 10^{-13} & 6.72\times 10^{-14} & -1.11\times 10^{-12} & 5.02\times 10^{-11} & 2.10\times 10^{-10} & -3.49\times 10^{-14} & 7.66\times 10^{-12} &
   2.19\times 10^{-11} \\
 -9.15\times 10^{-16} & 7.35\times 10^{-16} & 2.91\times 10^{-14} & 7.10\times 10^{-15} & -8.03\times 10^{-15} & -8.12\times 10^{-14} & -3.49\times 10^{-14} & 3.22\times 10^{-13} & 8.29\times 10^{-13} &
   -1.11\times 10^{-12} \\
 2.07\times 10^{-13} & -2.15\times 10^{-14} & -1.31\times 10^{-12} & 4.64\times 10^{-13} & -8.12\times 10^{-14} & 2.10\times 10^{-12} & 7.66\times 10^{-12} & 8.29\times 10^{-13} & 2.44\times 10^{-11} &
   5.02\times 10^{-11} \\
 5.75\times 10^{-13} & -6.72\times 10^{-14} & -5.51\times 10^{-12} & 4.30\times 10^{-13} & -3.49\times 10^{-14} & 7.66\times 10^{-12} & 2.19\times 10^{-11} & -1.11\times 10^{-12} & 5.02\times 10^{-11} &
   2.10\times 10^{-10} \\
\end{array}
\right)
\end{equation}

\subsection*{D meson meson decays}
\normalsize

\noindent The bounds of table \ref{table:exptal bounds} on leptonic D meson decay give the following covariance matrix in the basis $(\epsilon_{P,L}^{e \mu cu},\epsilon_{A,L}^{e \mu cu} , \epsilon_{P,R}^{e \mu cu},\epsilon_{A,R}^{e \mu cu})$ : 

\begin{equation}
\left(
\begin{array}{cccc}
 3.07\times 10^{-6} & -3.55 \times10^{-7} & -2.86\times 10^{-8} & 7.91\times 10^{-5} \\
 -3.55\times 10^{-7} & 2.04 \times10^{-3} & 7.91\times 10^{-5} & 7.30\times 10^{-7} \\
 -2.86\times 10^{-8} & 7.91\times 10^{-5} & 3.07\times 10^{-6} & -3.55\times 10^{-7} \\
 7.91\times 10^{-5} & 7.30\times 10^{-7} & -3.55\times 10^{-7} & 2.04\times 10^{-3} \\
\end{array}
\right)
\end{equation} \\

\noindent Using bounds on the semileptonic decay of D and $D_s$ meson give  in the basis $(\epsilon_{S,L}^{e \mu cu} ,\epsilon_{V,L}^{e \mu cu},\epsilon_{T_L}^{e \mu cu},\epsilon_{S,R}^{e \mu cu},\epsilon_{V,R}^{e \mu cu},\epsilon_{T_R}^{e \mu cu})$ : \\ \\

\begin{equation}
\hspace{-1.5cm}
\left(
\begin{array}{cccccc}
 1.80\times 10^{-6} & 1.32\times 10^{-7} & -3.19\times 10^{-8} & -2.10\times 10^{-8} & -1.61\times 10^{-7} & 1.79\times 10^{-8} \\
 1.32\times 10^{-7} & 2.10\times 10^{-6} & 3.65\times 10^{-7} & -1.61\times 10^{-7} & 9.7\times 10^{-8} & 7.06\times 10^{-7} \\
 -3.19\times 10^{-8} & 3.65\times 10^{-7} & 4.03\times 10^{-6} & 1.79\times 10^{-8} & 7.06\times 10^{-7} & 2.30\times 10^{-7} \\
 -2.10\times 10^{-8} & -1.61\times 10^{-7} & 1.79\times 10^{-8} & 1.80\times 10^{-6} & 1.32\times 10^{-7} & -3.19\times 10^{-8} \\
 -1.61\times 10^{-7} & 9.7\times 10^{-8} & 7.06\times 10^{-7} & 1.32\times 10^{-7} & 2.10\times 10^{-6} & 3.65\times 10^{-7} \\
 1.79\times 10^{-8} & 7.06\times 10^{-7} & 2.30\times 10^{-7} & -3.19\times 10^{-8} & 3.65\times 10^{-7} & 4.03\times 10^{-6} \\
\end{array}
\right)
\end{equation}

\noindent The covariance matrix in the basis $\left( \begin{array}{c}\epsilon_{P,L}^{e \mu cu},\epsilon_{A,L}^{e \mu cu},\epsilon_{P,R}^{e \mu cu},\epsilon_{A,R}^{e \mu cu},\epsilon_{S,L}^{e \mu cu},\epsilon_{V,L}^{e \mu cu},\epsilon_{T_{L}}^{e \mu cu}, \epsilon_{S,R}^{e \mu cu},\epsilon_{V,R}^{e \mu cu},\epsilon_{T_{R}}^{e \mu cu}\\\end{array} \right)_{\Lambda_W}$ is :

\tiny

\begin{equation}
\hspace{-2.5cm}
\left(
\begin{array}{cccccccccc}
 1.15 \times 10^{-6} & -2.16 \times 10^{-7} & -1.15 \times 10^{-8} & 4.81 \times 10^{-5} & -1.45 \times 10^{-8} & -2.62 \times 10^{-8} & -2.97 \times 10^{-7} & -1.55 \times 10^{-9} & -8.69 \times 10^{-7} & -1.68 \times 10^{-8} \\
 -2.16 \times 10^{-7} & 2.04 \times 10^{-3} & 4.81 \times 10^{-5} & 7.31 \times 10^{-7} & 1.81 \times 10^{-9} & 3.50 \times 10^{-5} & 8.22 \times 10^{-9} & 8.70 \times 10^{-9} & -1.09 \times 10^{-8} & 1.99 \times 10^{-7} \\
 -1.15 \times 10^{-8} & 4.81 \times 10^{-5} & 1.15 \times 10^{-6} & -2.16 \times 10^{-7} & 1.55 \times 10^{-9} & 8.69 \times 10^{-7} & 1.68 \times 10^{-8} & 1.45 \times 10^{-8} & 2.62 \times 10^{-8} & 2.97 \times 10^{-7} \\
 4.81 \times 10^{-5} & 7.31 \times 10^{-7} & -2.16 \times 10^{-7} & 2.04 \times 10^{-3} & -8.70 \times 10^{-9} & 1.09 \times 10^{-8} & -1.99 \times 10^{-7} & -1.81 \times 10^{-9} & -3.50 \times 10^{-5} & -8.22 \times 10^{-9} \\
 -1.45 \times 10^{-8} & 1.81 \times 10^{-9} & 1.55 \times 10^{-9} & -8.70 \times 10^{-9} & 6.80 \times 10^{-7} & 1.03 \times 10^{-7} & 2.73 \times 10^{-7} & -5.58 \times 10^{-9} & -5.42 \times 10^{-8} & 2.96 \times 10^{-8} \\
 -2.62 \times 10^{-8} & 3.50 \times 10^{-5} & 8.69 \times 10^{-7} & 1.09 \times 10^{-8} & 1.03 \times 10^{-7} & 2.70 \times 10^{-6} & 4.31 \times 10^{-7} & -5.42 \times 10^{-8} & 9.66 \times 10^{-8} & 8.36 \times 10^{-7} \\
 -2.97 \times 10^{-7} & 8.22 \times 10^{-9} & 1.68 \times 10^{-8} & -1.99 \times 10^{-7} & 2.73 \times 10^{-7} & 4.31 \times 10^{-7} & 5.62 \times 10^{-6} & 2.96 \times 10^{-8} & 8.36 \times 10^{-7} & 3.21 \times 10^{-7} \\
 -1.55 \times 10^{-9} & 8.70 \times 10^{-9} & 1.45 \times 10^{-8} & -1.81 \times 10^{-9} & -5.58 \times 10^{-9} & -5.42 \times 10^{-8} & 2.96 \times 10^{-8} & 6.80 \times 10^{-7} & 1.03 \times 10^{-7} & 2.73 \times 10^{-7} \\
 -8.69 \times 10^{-7} & -1.09 \times 10^{-8} & 2.62 \times 10^{-8} & -3.50 \times 10^{-5} & -5.42 \times 10^{-8} & 9.66 \times 10^{-8} & 8.36 \times 10^{-7} & 1.03 \times 10^{-7} & 2.70 \times 10^{-6} & 4.31 \times 10^{-7} \\
 -1.68 \times 10^{-8} & 1.99 \times 10^{-7} & 2.97 \times 10^{-7} & -8.22 \times 10^{-9} & 2.96 \times 10^{-8} & 8.36 \times 10^{-7} & 3.21 \times 10^{-7} & 2.73 \times 10^{-7} & 4.31 \times 10^{-7} & 5.62 \times 10^{-6} \\
\end{array}
\right)
\end{equation}

\subsection*{B meson decays}
\normalsize
\noindent The bound on the leptonic decay of the B meson (see table \ref{table:exptal bounds}) gives the following covariance matrix in the basis $(\epsilon_{P,L}^{e \mu bd},\epsilon_{A,L}^{e \mu bd},\epsilon_{P,R}^{e \mu bd},\epsilon_{A,R}^{e \mu bd})$   : \\

\begin{equation}
\left(
\begin{array}{cccc}
 5.53 \times 10^{-8} & 9.23 \times 10^{-8} & 1.20 \times 10^{-9} & 3.48 \times 10^{-6} \\
 9.23 \times 10^{-8} & 2.20 \times 10^{-4} & 3.48 \times 10^{-6} & 6.89 \times 10^{-6} \\
 1.20 \times 10^{-9} & 3.48 \times 10^{-6} & 5.53 \times 10^{-8} & 9.23 \times 10^{-8} \\
 3.48 \times 10^{-6} & 6.89 \times 10^{-6} & 9.23 \times 10^{-8} & 2.20 \times 10^{-4} \\
\end{array}
\right)
\end{equation} \\

\noindent The covariance matrix in the basis   $(\epsilon_{S,L}^{e \mu bd},\epsilon_{V,L}^{e \mu bd}, \epsilon_{T_L}^{e \mu bd},\epsilon_{S,R}^{e \mu bd},\epsilon_{V,R}^{e \mu bd},\epsilon_{T_R}^{e \mu bd})$ is : \\ \\ 

\begin{equation}
\hspace{-1.8cm}
\left(
\begin{array}{cccccc}
 2.07\times 10^{-10} & 1.21\times 10^{-11} & 1.52\times 10^{-12} & -3.90\times 10^{-15} & -5.74\times 10^{-14} & 5.18\times 10^{-15} \\
 1.21\times 10^{-11} & 2.23\times 10^{-10} & 2.81\times 10^{-11} & -5.74\times 10^{-14} & 2.87\times 10^{-14} & 2.32\times 10^{-13} \\
 1.52\times 10^{-12} & 2.81\times 10^{-11} & 4.03\times 10^{-10} & 5.18\times 10^{-15} & 2.32\times 10^{-13} & 3.50\times 10^{-14} \\
 -3.90\times 10^{-15} & -5.74\times 10^{-14} & 5.18\times 10^{-15} & 2.07\times 10^{-10} & 1.21\times 10^{-11} & 1.52\times 10^{-12} \\
 -5.74\times 10^{-14} & 2.87\times 10^{-14} & 2.32\times 10^{-13} & 1.21\times 10^{-11} & 2.23\times 10^{-10} & 2.81\times 10^{-11} \\
 5.18\times 10^{-15} & 2.32\times 10^{-13} & 3.50\times 10^{-14} & 1.52\times 10^{-12} & 2.81\times 10^{-11} & 4.03\times 10^{-10} \\
\end{array}
\right)
\end{equation}

\normalsize
\noindent The covariance matrix in the basis $\left( \begin{array}{c}\epsilon_{P,L}^{e \mu bd},\epsilon_{A,L}^{e \mu bd},\epsilon_{P,R}^{e \mu bd},\epsilon_{A,R}^{e \mu bd},\epsilon_{S,L}^{e \mu bd},\epsilon_{V,L}^{e \mu bd},\epsilon_{T_{L}}^{e \mu bd}, \epsilon_{S,R}^{e \mu bd},\epsilon_{V,R}^{e \mu bd},\epsilon_{T_{R}}^{e \mu bd}\\\end{array} \right)_{\Lambda_W}$ is :

\tiny

\begin{equation}
\hspace{-2.5cm}
\left(
\begin{array}{cccccccccc}
 2.74 \times 10^{-8} & 6.51 \times 10^{-8} & 5.94 \times 10^{-10} & 2.45 \times 10^{-6} & -1.10 \times 10^{-12} & -4.46 \times 10^{-10} & 5.02 \times 10^{-11} & 1.89 \times 10^{-14} & 1.68 \times 10^{-8} & -8.41 \times 10^{-13} \\
 6.51 \times 10^{-8} & 2.20 \times 10^{-4} & 2.45 \times 10^{-6} & 6.89 \times 10^{-6} & -2.11 \times 10^{-12} & -1.51 \times 10^{-6} & 9.19 \times 10^{-11} & 7.76 \times 10^{-11} & 4.73 \times 10^{-8} & -3.47 \times 10^{-9} \\
 5.94 \times 10^{-10} & 2.45 \times 10^{-6} & 2.74 \times 10^{-8} & 6.51 \times 10^{-8} & -1.89 \times 10^{-14} & -1.68 \times 10^{-8} & 8.41 \times 10^{-13} & 1.10 \times 10^{-12} & 4.46 \times 10^{-10} & -5.02 \times 10^{-11} \\
 2.45 \times 10^{-6} & 6.89 \times 10^{-6} & 6.51 \times 10^{-8} & 2.20 \times 10^{-4} & -7.76 \times 10^{-11} & -4.73 \times 10^{-8} & 3.47 \times 10^{-9} & 2.11 \times 10^{-12} & 1.51 \times 10^{-6} & -9.19 \times 10^{-11} \\
 -1.10 \times 10^{-12} & -2.11 \times 10^{-12} & -1.89 \times 10^{-14} & -7.76 \times 10^{-11} & 1.03 \times 10^{-10} & 7.83 \times 10^{-12} & -1.03 \times 10^{-11} & -2.10 \times 10^{-15} & -5.78 \times 10^{-13} & 3.15 \times 10^{-15} \\
 -4.46 \times 10^{-10} & -1.51 \times 10^{-6} & -1.68 \times 10^{-8} & -4.73 \times 10^{-8} & 7.83 \times 10^{-12} & 1.06 \times 10^{-8} & 3.09 \times 10^{-11} & -5.78 \times 10^{-13} & -3.24 \times 10^{-10} & 2.41 \times 10^{-11} \\
 5.02 \times 10^{-11} & 9.19 \times 10^{-11} & 8.41 \times 10^{-13} & 3.47 \times 10^{-9} & -1.03 \times 10^{-11} & 3.09 \times 10^{-11} & 5.10 \times 10^{-10} & 3.15 \times 10^{-15} & 2.41 \times 10^{-11} & 4.30 \times 10^{-14} \\
 1.89 \times 10^{-14} & 7.76 \times 10^{-11} & 1.10 \times 10^{-12} & 2.11 \times 10^{-12} & -2.10 \times 10^{-15} & -5.78 \times 10^{-13} & 3.15 \times 10^{-15} & 1.03 \times 10^{-10} & 7.83 \times 10^{-12} & -1.03 \times 10^{-11} \\
 1.68 \times 10^{-8} & 4.73 \times 10^{-8} & 4.46 \times 10^{-10} & 1.51 \times 10^{-6} & -5.78 \times 10^{-13} & -3.24 \times 10^{-10} & 2.41 \times 10^{-11} & 7.83 \times 10^{-12} & 1.06 \times 10^{-8} & 3.09 \times 10^{-11} \\
 -8.41 \times 10^{-13} & -3.47 \times 10^{-9} & -5.02 \times 10^{-11} & -9.19 \times 10^{-11} & 3.15 \times 10^{-15} & 2.41 \times 10^{-11} & 4.30 \times 10^{-14} & -1.03 \times 10^{-11} & 3.09 \times 10^{-11} & 5.10 \times 10^{-10} \\
\end{array}
\right)
\end{equation}

\subsection*{Bs meson}
\normalsize
\noindent The bound on the leptonic decay of the $B_s$ meson gives in the basis  $(\epsilon_{P,L}^{e \mu bs},\epsilon_{A,L}^{e \mu bs},\epsilon_{P,R}^{e \mu bs},\epsilon_{A,R}^{e \mu bs})$ : \\ 

\begin{equation}
\left(
\begin{array}{cccc}
 3.06 \times 10^{-8} & -1.22 \times 10^{-8} & -3.40 \times 10^{-10} & 1.94 \times 10^{-6} \\
 -1.22 \times 10^{-8} & 1.24 \times 10^{-4} & 1.94 \times 10^{-6} & -1.80 \times 10^{-7} \\
 -3.40 \times 10^{-10} & 1.94 \times 10^{-6} & 3.06 \times 10^{-8} & -1.22 \times 10^{-8} \\
 1.94 \times 10^{-6} & -1.80 \times 10^{-7} & -1.22 \times 10^{-8} & 1.24 \times 10^{-4} \\
\end{array}
\right)
\end{equation} \\ \\

\noindent The bound on the $B_s$ meson  decaying into Kaon (table \ref{table:exptal bounds}) gives in the basis  $(\epsilon_{S,L}^{e \mu bs},\epsilon_{V,L}^{e \mu bs}, \epsilon_{T_L}^{e \mu bs},\epsilon_{S,R}^{e \mu bs},\epsilon_{V,R}^{e \mu bs},\epsilon_{T_R}^{e \mu bs})$  : \\ \\

\begin{equation}
\hspace{-2cm}
\left(
\begin{array}{cccccc}
 5.05 \times 10^{-10} & 3.47 \times 10^{-11} & 5.07 \times 10^{-12} & -1.13 \times 10^{-14} & -1.65 \times 10^{-13} & 1.73 \times 10^{-14} \\
 3.47 \times 10^{-11} & 6.53 \times 10^{-10} & 9.54 \times 10^{-11} & -1.65 \times 10^{-13} & 8.78 \times 10^{-14} & 7.90 \times 10^{-13} \\
 5.07 \times 10^{-12} & 9.54 \times 10^{-11} & 1.51 \times 10^{-9} & 1.73 \times 10^{-14} & 7.90 \times 10^{-13} & 1.38 \times 10^{-13} \\
 -1.13 \times 10^{-14} & -1.65 \times 10^{-13} & 1.73 \times 10^{-14} & 5.05 \times 10^{-10} & 3.47 \times 10^{-11} & 5.07 \times 10^{-12} \\
 -1.65 \times 10^{-13} & 8.78 \times 10^{-14} & 7.90 \times 10^{-13} & 3.47 \times 10^{-11} & 6.53 \times 10^{-10} & 9.54 \times 10^{-11} \\
 1.73 \times 10^{-14} & 7.90 \times 10^{-13} & 1.38 \times 10^{-13} & 5.07 \times 10^{-12} & 9.54 \times 10^{-11} & 1.51 \times 10^{-9} \\
\end{array}
\right)
\end{equation}

\normalsize
\noindent The covariance matrix in the basis $\left( \begin{array}{c}\epsilon_{P,L}^{e \mu bs},\epsilon_{A,L}^{e \mu bs},\epsilon_{P,R}^{e \mu bs},\epsilon_{A,R}^{e \mu bs},\epsilon_{S,L}^{e \mu bs},\epsilon_{V,L}^{e \mu bs},\epsilon_{T_{L}}^{e \mu bs}, \epsilon_{S,R}^{e \mu bs},\epsilon_{V,R}^{e \mu bs},\epsilon_{T_{R}}^{e \mu bs}\\\end{array} \right)_{\Lambda_W}$ is :

\tiny

\begin{equation}
\hspace{-2.5cm}
\left(
\begin{array}{cccccccccc}
 1.52 \times 10^{-8} & -8.62 \times 10^{-9} & -1.69 \times 10^{-10} & 1.37 \times 10^{-6} & -1.35 \times 10^{-12} & 6.16 \times 10^{-11} & 6.41 \times 10^{-11} & -5.11 \times 10^{-15} & 9.39 \times 10^{-9} & 2.42 \times 10^{-13} \\
 -8.62 \times 10^{-9} & 1.24 \times 10^{-4} & 1.37 \times 10^{-6} & -1.80 \times 10^{-7} & 1.21 \times 10^{-13} & -8.51 \times 10^{-7} & -1.29 \times 10^{-11} & 4.33 \times 10^{-11} & -1.24 \times 10^{-9} & -1.94 \times 10^{-9} \\
 -1.69 \times 10^{-10} & 1.37 \times 10^{-6} & 1.52 \times 10^{-8} & -8.62 \times 10^{-9} & 5.11 \times 10^{-15} & -9.39 \times 10^{-9} & -2.42 \times 10^{-13} & 1.35 \times 10^{-12} & -6.16 \times 10^{-11} & -6.41 \times 10^{-11} \\
 1.37 \times 10^{-6} & -1.80 \times 10^{-7} & -8.62 \times 10^{-9} & 1.24 \times 10^{-4} & -4.33 \times 10^{-11} & 1.24 \times 10^{-9} & 1.94 \times 10^{-9} & -1.21 \times 10^{-13} & 8.51 \times 10^{-7} & 1.29 \times 10^{-11} \\
 -1.35 \times 10^{-12} & 1.21 \times 10^{-13} & 5.11 \times 10^{-15} & -4.33 \times 10^{-11} & 2.51 \times 10^{-10} & 2.21 \times 10^{-11} & -3.90 \times 10^{-11} & -6.11 \times 10^{-15} & -4.33 \times 10^{-13} & 9.78 \times 10^{-15} \\
 6.16 \times 10^{-11} & -8.51 \times 10^{-7} & -9.39 \times 10^{-9} & 1.24 \times 10^{-9} & 2.21 \times 10^{-11} & 6.49 \times 10^{-9} & 1.07 \times 10^{-10} & -4.33 \times 10^{-13} & 8.57 \times 10^{-12} & 1.42 \times 10^{-11} \\
 6.41 \times 10^{-11} & -1.29 \times 10^{-11} & -2.42 \times 10^{-13} & 1.94 \times 10^{-9} & -3.90 \times 10^{-11} & 1.07 \times 10^{-10} & 1.91 \times 10^{-9} & 9.78 \times 10^{-15} & 1.42 \times 10^{-11} & 1.74 \times 10^{-13} \\
 -5.11 \times 10^{-15} & 4.33 \times 10^{-11} & 1.35 \times 10^{-12} & -1.21 \times 10^{-13} & -6.11 \times 10^{-15} & -4.33 \times 10^{-13} & 9.78 \times 10^{-15} & 2.51 \times 10^{-10} & 2.21 \times 10^{-11} & -3.90 \times 10^{-11} \\
 9.39 \times 10^{-9} & -1.24 \times 10^{-9} & -6.16 \times 10^{-11} & 8.51 \times 10^{-7} & -4.33 \times 10^{-13} & 8.57 \times 10^{-12} & 1.42 \times 10^{-11} & 2.21 \times 10^{-11} & 6.49 \times 10^{-9} & 1.07 \times 10^{-10} \\
 2.42 \times 10^{-13} & -1.94 \times 10^{-9} & -6.41 \times 10^{-11} & 1.29 \times 10^{-11} & 9.78 \times 10^{-15} & 1.42 \times 10^{-11} & 1.74 \times 10^{-13} & -3.90 \times 10^{-11} & 1.07 \times 10^{-10} & 1.91 \times 10^{-9} \\
\end{array}
\right)
\end{equation} \\

\end{appendices}

\normalsize
\newpage

\bibliographystyle{ieeetr} 
\bibliography{meson4}

\begin{thebibliography}{10}

\bibitem{numassesoscill}
S.-K. Collaboration, Y.~Fukuda, and al, ``Evidence for oscillation of
  atmospheric neutrinos,'' {\em DOI:10.1103/PhysRevLett.81.1562}.
\newblock \href{https://arxiv.org/abs/hep-ex/9807003}{[arXiv:hep-ex/9807003]}.

\bibitem{neutrinoosci2}
Q.~R. Ahmad {\em et~al.}, ``{Direct evidence for neutrino flavor transformation
  from neutral current interactions in the Sudbury Neutrino Observatory},''
  {\em Phys. Rev. Lett.}, vol.~89, p.~011301, 2002.

\bibitem{PDG}
K.~Olive and al, ``Particle data group,'' {\em Chin.Phys. C,38, 090001, 2017
  update}.

\bibitem{LHC1}
{CMS Collaboration}, ``Search for heavy majorana neutrinos in e+/- e+/- plus
  jets and e+/- mu+/- plus jets events in proton-proton collisions at sqrt(s) =
  8 tev,'' {\em DOI:10.1007/JHEP04(2016)169}.
\newblock \href{https://arxiv.org/abs/1603.02248}{[arXiv:1603.02248 [hep-ex]]}.

\bibitem{LHC2}
{ATLAS Collaboration}, ``Search for heavy majorana neutrinos with the atlas
  detector in pp collisions at s√=8 tev,'' {\em DOI:10.1007/JHEP07(2015)162}.
\newblock \href{https://arxiv.org/abs/1506.06020}{[arXiv:1506.06020 [hep-ex]]}.

\bibitem{lowLFV3}
Y.~Kuno and Y.~Okada, ``Muon decay and physics beyond the standard model,''
  {\em DOI:10.1103/RevModPhys.73.151}.
\newblock \href{https://arxiv.org/abs/hep-ph/9909265}{[arXiv:hep-ph/9909265]}.

\bibitem{CLFVintrol}
L.~Calibbi and G.~Signorelli, ``Charged lepton flavour violation: An
  experimental and theoretical introduction,'' {\em
  DOI:10.1393/ncr/i2018-10144-0}.
\newblock \href{https://arxiv.org/abs/1709.00294}{[arXiv:1709.00294[hep-ph]]}.

\bibitem{Arganda:2004bz}
E.~Arganda, A.~M. Curiel, M.~J. Herrero, and D.~Temes, ``{Lepton flavor
  violating Higgs boson decays from massive seesaw neutrinos},'' {\em Phys.
  Rev.}, vol.~D71, p.~035011, 2005.

\bibitem{Raidal:2008jk}
M.~Raidal {\em et~al.}, ``{Flavour physics of leptons and dipole moments},''
  {\em Eur. Phys. J.}, vol.~C57, pp.~13--182, 2008.

\bibitem{MEG}
A.~Baldini and al. [MEG~Collaboration], ``Search for the lepton flavour
  violating decay $ \mu^+ \rightarrow e^+ \gamma$ with the full dataset of the
  meg experiment,'' {\em Eur. Phys. J. C 76 (2016) no.8, 434
  doi:10.1140/epjc/s10052-016-4271-x}.
\newblock \href{https://arxiv.org/abs/1605.05081}{[arXiv:1605.05081 [hep-ex]]}.

\bibitem{SINDRUM}
U.~B. et~al. [SINDRUM~Collaboration], ``Search for the decay µ → 3e,'' {\em
  Nucl. Phys. B 299 (1988) 1. doi:10.1016/0550-3213(88)90462-2}.

\bibitem{mu3e}
{A. Perrevoort [Mu3e Collaboration]}, ``{ Status of the Mu3e Experiment at
  PSI},'' {\em DOI:10.1051/epjconf/201611801028}.
\newblock \href{https://arxiv.org/abs/1605.02906}{[arXiv:1605.02906v1
  [physics.ins-det]]}.

\bibitem{SINDRUM2}
W.~H.~B. et~al. [SINDRUM II~Collaboration], ``A search for muon to electron
  conversion in muonic gold,'' {\em Eur. Phys. J. C 47 (2006) 337.
  doi:10.1140/epjc/s2006-02582-x C}.

\bibitem{Comet}
{Y. Kuno [COMET Collaboration]}, ``{A search for muon-to-electron conversion at
  J-PARC: The COMET experiment},'' {\em PTEP 2013 (2013) 022C01.
  doi:10.1093/ptep/pts089}.

\bibitem{Mu2e}
{R. M. Carey et al. [Mu2e Collaboration]}, ``{Proposal to search for $\mu^-N
  \rightarrow e^-N$ with a single event sensitivity below $10^{-16}$},'' {\em
  FERMILAB-PROPOSAL-0973}.

\bibitem{K0L}
{D. Ambrose et al, BNL E871 Collaboration}, ``{New Limit on Muon and Electron
  Lepton Number Violation from
  ${\mathit{K}}_{\mathit{L}}^{0}\ensuremath{\rightarrow}{\mathit{\ensuremath{\mu}}}^{\ifmmode\pm\else\textpm\fi{}}{\mathit{e}}^{\ensuremath{\mp}}$
  Decay},'' {\em Phys. Rev. Lett.}, vol.~81, pp.~5734--5737, Dec 1998.

\bibitem{D0}
{ LHCb collaboration and R. Aaij }, ``{Search for the lepton-flavour violating
  decay $D^0 \rightarrow e^{\pm} \mu^{\mp}$},'' {\em
  DOI:10.1016/j.physletb.2016.01.029}.
\newblock \href{https://arxiv.org/abs/1512.00322}{[arXiv:1512.00322v2
  [hep-ex]]}.

\bibitem{B0S}
{R.Aaij, LHCb Collaboration}, ``Search for the lepton-flavor-violating decays
  ${B}_{s}^{0}\ensuremath{\rightarrow}{e}^{\mathbf{\ifmmode\pm\else\textpm\fi{}}}{\ensuremath{\mu}}^{\mathbf{\ensuremath{\mp}}}$
  and
  ${B}^{0}\ensuremath{\rightarrow}{e}^{\mathbf{\ifmmode\pm\else\textpm\fi{}}}{\ensuremath{\mu}}^{\mathbf{\ensuremath{\mp}}}$,''
  {\em Phys. Rev. Lett.}, vol.~111, p.~141801, Sep 2013.

\bibitem{K+e+}
{R. Appel et al}, ``{Search for Lepton Flavor Violation in $K^+$ Decays},''
  {\em DOI:10.1103/PhysRevLett.85.2877}.
\newblock
  \href{https://arxiv.org/abs/hep-ex/0006003}{[arXiv:hep-ex/0006003v1]}.

\bibitem{D+}
{ The BABAR Collaboration and J. P. Lees }, ``Searches for rare or forbidden
  semileptonic charm decays,'' {\em DOI:10.1103/PhysRevD.84.072006}.
\newblock \href{https://arxiv.org/abs/1107.4465}{[arXiv:1107.4465v1 [hep-ex]]}.

\bibitem{Bpi+}
{ The BABAR Collaboration and B. Aubert}, ``{Search for the rare decay B to pi
  l+ l-},'' {\em DOI:10.1103/PhysRevLett.99.051801}.
\newblock
  \href{https://arxiv.org/abs/hep-ex/0703018v2}{[arXiv:hep-ex/0703018]]}.

\bibitem{BK+}
{ The BABAR Collaboration and B. Aubert}, ``{Measurements of branching
  fractions, rate asymmetries, and angular distributions in the rare decays B
  --> Kl+l- and B --> K*l+l-},'' {\em DOI:10.1103/PhysRevD.73.092001}.
\newblock
  \href{https://arxiv.org/abs/hep-ex/0604007v2}{[arXiv:hep-ex/0604007v2]]}.

\bibitem{EFTgeorgi1}
H.~Georgi, ``Effective field theory,'' {\em Ann. Rev. Nucl. Part. Sci. 43
  (1993) 209-252}.

\bibitem{Manohar:2018aog}
A.~V. Manohar, ``{Introduction to Effective Field Theories},'' in {\em {Les
  Houches summer school: EFT in Particle Physics and Cosmology Les Houches,
  Chamonix Valley, France, July 3-28, 2017}}, 2018.

\bibitem{Pich:2018ltt}
A.~Pich, ``{Effective Field Theory with Nambu-Goldstone Modes},'' in {\em {Les
  Houches summer school: EFT in Particle Physics and Cosmology Les Houches,
  Chamonix Valley, France, July 3-28, 2017}}, 2018.

\bibitem{LQ}
S.~Davidson, D.~C. Bailey, and B.~A. Campbell, ``{Model independent constraints
  on leptoquarks from rare processes},'' {\em Z. Phys.}, vol.~C61,
  pp.~613--644, 1994.

\bibitem{tau}
D.~Black, T.~Han, H.-J. He, and M.~Sher, ``{tau - mu flavor violation as a
  probe of the scale of new physics},'' {\em Phys. Rev.}, vol.~D66, p.~053002,
  2002.

\bibitem{Carpentier}
M.~Carpentier and S.~Davidson, ``Constraints on two-lepton, two quark
  operators,'' {\em doi:10.1140/epjc/s10052-010-1482-4}.
\newblock \href{https://arxiv.org/abs/1008.0280}{[arXiv:1008.0280[hep-ph]]}.

\bibitem{Cai:2015poa}
Y.~Cai and M.~A. Schmidt, ``{A Case Study of the Sensitivity to LFV Operators
  with Precision Measurements and the LHC},'' {\em JHEP}, vol.~02, p.~176,
  2016.

\bibitem{Petrov16}
D.~E. Hazard and A.~A. Petrov, ``{Lepton flavor violating quarkonium decays},''
  {\em Phys. Rev.}, vol.~D94, no.~7, p.~074023, 2016.

\bibitem{Petrov17}
D.~E. Hazard and A.~A. Petrov, ``{Radiative lepton flavor violating B, D, and K
  decays},'' 2017.

\bibitem{Buras}
G.~Buchalla, A.~Buras, and M.~E. Lautenbacher, ``Weak decays beyond leading
  logarithms,'' {\em DOI:10.1103/RevModPhys.68.1125}.
\newblock \href{https://arxiv.org/abs/hep-ph/9512380}{[arXiv:hep-ph/9512380]}.

\bibitem{LHCb1}
R.~Aaij {\em et~al.}, ``{Test of lepton universality using $B^{+}\rightarrow
  K^{+}\ell^{+}\ell^{-}$ decays},'' {\em Phys. Rev. Lett.}, vol.~113,
  p.~151601, 2014.

\bibitem{LHCb2}
R.~Aaij {\em et~al.}, ``{Test of lepton universality with $B^{0} \rightarrow
  K^{*0}\ell^{+}\ell^{-}$ decays},'' {\em JHEP}, vol.~08, p.~055, 2017.

\bibitem{LHCb}
{[LHCb collaboration]}, ``{Measurement of the ratio of branching fractions
  $\mathcal{B}$($\bar{B}^0$ $\rightarrow$
  $D^{*+}\tau^{-}\bar{\nu}_{\tau}$/$\mathcal{B}$($\bar{B}^0$ $\rightarrow$
  $D^{*+}\mu^{-}\bar{\nu}_{\mu}$)},'' {\em DOI:10.1103/PhysRevLett.115.111803}.
\newblock \href{https://arxiv.org/abs/1506.08614}{[arXiv:1506.08614v2
  [hep-ex]]}.

\bibitem{belle}
{Belle Collaboration, Y. Sato et al}, ``{Measurement of the branching ratio of
  $\bar{B}^0$ $\rightarrow$ $D^{*+}\tau^{-}\bar{\nu}_{\tau}$ relative to
  $\bar{B}^0$ $\rightarrow$ $D^{*+}l\bar{\nu}_{l}$ decays with a semileptonic
  tagging method},'' {\em DOI:10.1103/PhysRevD.94.072007}.
\newblock \href{https://arxiv.org/pdf/1607.07923.pdf}{[arXiv:1607.07923v3
  [hep-ex]]}.

\bibitem{babar}
{BaBar Collaboration, J. P. Lees et al}, ``{Measurement of an Excess of
  $\bar{B}$ $\rightarrow$ $D^{(*)}\tau^{-}\bar{\nu}_{\tau}$ Decays and
  Implications for Charged Higgs Bosons},'' {\em
  DOI:10.1103/PhysRevD.88.072012}.
\newblock \href{https://arxiv.org/abs/1303.0571}{[arXiv:1303.0571v1 [hep-ex]]}.

\bibitem{SDG}
S.~Descotes-Genon, L.~Hofer, J.~Matias, and J.~Virto, ``{Global analysis of
  $b\to s\ell\ell$ anomalies},'' {\em JHEP}, vol.~06, p.~092, 2016.

\bibitem{Luca}
M.~Ciuchini, M.~Fedele, E.~Franco, S.~Mishima, A.~Paul, L.~Silvestrini, and
  M.~Valli, ``{$B\to K^* \ell^+ \ell^-$ decays at large recoil in the Standard
  Model: a theoretical reappraisal},'' {\em JHEP}, vol.~06, p.~116, 2016.

\bibitem{SJ14}
S.~Jäger and J.~Martin~Camalich, ``{Reassessing the discovery potential of the
  $B \to K^{*} \ell^+\ell^-$ decays in the large-recoil region: SM challenges
  and BSM opportunities},'' {\em Phys. Rev.}, vol.~D93, no.~1, p.~014028, 2016.

\bibitem{SJ17}
L.-S. Geng, B.~Grinstein, S.~Jäger, J.~Martin~Camalich, X.-L. Ren, and R.-X.
  Shi, ``{Towards the discovery of new physics with lepton-universality ratios
  of $b\to s\ell\ell$ decays},'' {\em Phys. Rev.}, vol.~D96, no.~9, p.~093006,
  2017.

\bibitem{Gundrun1}
G.~Hiller and M.~Schmaltz, ``{$R_K$ and future $b \to s \ell \ell$ physics
  beyond the standard model opportunities},'' {\em Phys. Rev.}, vol.~D90,
  p.~054014, 2014.

\bibitem{Bobeth}
C.~Bobeth, G.~Hiller, and D.~van Dyk, ``{General analysis of $\bar{B} \to
  \bar{K}^{(*)}\ell^+ \ell^-$ decays at low recoil},'' {\em Phys. Rev.},
  vol.~D87, no.~3, p.~034016, 2013.
\newblock [Phys. Rev.D87,034016(2013)].

\bibitem{APS}
W.~Altmannshofer, P.~Paradisi, and D.~M. Straub, ``{Model-Independent
  Constraints on New Physics in $b \to s$ Transitions},'' {\em JHEP}, vol.~04,
  p.~008, 2012.

\bibitem{MLM}
S.~Davidson, M.~L. Mangano, S.~Perries, and V.~Sordini, ``{Lepton Flavour
  Violating top decays at the LHC},'' {\em Eur. Phys. J.}, vol.~C75, no.~9,
  p.~450, 2015.

\bibitem{RGE1}
S.~Davidson, ``Mu to e gamma and matching at mw,'' {\em
  doi:10.1140/epjc/s10052-016-4207-5}.
\newblock \href{https://arxiv.org/abs/1601.07166}{[arXiv:1601.07166 [hep-ph]]}.

\bibitem{RGE2}
A.~Crivellin, S.~Davidson, G.~M. Pruna, and A.~Signer, ``Renormalisation-group
  improved analysis of µ → e processes in a systematic
  effective-field-theory approach,'' {\em doi:10.1007/JHEP05(2017)117}.
\newblock \href{https://arxiv.org/abs/1702.03020}{[arXiv:1702.03020 [hep-ph]]}.

\bibitem{Shanker}
{O.Shanker}, ``{Flavour violation, scalar particles and leptoquarks},'' {\em
  Nucl. Phys. B206 (1982) 253 DOI:10.1016/0550-3213(82)90534-X}.

\bibitem{Bonn}
M.~Herz, ``Bounds on leptoquark and supersymmetric,r-parity violating
  interactions from meson decays,''
\newblock \href{https://arxiv.org/abs/hep-ph/0301079}{[arXiv:hep-ph/0301079]}.

\bibitem{Khadroelemevol}
J.~Bijnens, G.~Colangelo, G.~Eckerand, and J.~Gasser, ``Semileptonic kaon
  decays,''
\newblock
  \href{https://arxiv.org/abs/hep-ph/9411311}{[arXiv:hep-ph/9411311v1]}.

\bibitem{Dhadroelemevol}
R.~Gupta, ``Calculations of hadronic matrix elements using lattice qcd,''
\newblock
  \href{https://arxiv.org/abs/hep-lat/9308002}{[arXiv:hep-lat/9308002]}.

\bibitem{Bhadroelemevol}
D.~Du, A.~X. El-Khadra, S.~Gottlieb, A.~S. Kronfeld, J.~Laiho, E.~Lunghi,
  R.~S.~V. de~Water, and R.~Zhou, ``Phenomenology of semileptonic b-meson
  decays with form factors from lattice qcd,'' {\em
  doi:10.1103/PhysRevD.93.034005}.
\newblock \href{https://arxiv.org/abs/1510.02349}{[arXiv:1510.02349 [hep-ph]]}.

\bibitem{semil1}
{Fermilab Lattice and MILC Collaborations}, ``{$|V_{ub}|$ from $B \to \pi l
  \nu$ decays and (2+1)-flavor lattice QCD},'' {\em
  DOI:10.1103/PhysRevD.92.014024}.
\newblock \href{https://arxiv.org/abs/1503.07839}{[arXiv:1503.07839v2
  [hep-lat]]}.

\bibitem{semil2}
{J. A. Bailey et al}, ``{$B\to Kl^+l^-$ decay form factors from three-flavor
  lattice QCD},'' {\em DOI:10.1103/PhysRevD.93.025026}.
\newblock \href{https://arxiv.org/abs/1509.06235v2}{[arXiv:1509.06235v2
  [hep-lat]]}.

\bibitem{semil3}
{P. Ball and R. Zwicky}, ``{New Results on B->pi, K, eta Decay Formfactors from
  Light-Cone Sum Rules },'' {\em DOI:10.1103/PhysRevD.71.014015}.
\newblock
  \href{https://arxiv.org/abs/hep-ph/0406232v1}{[arXiv:hep-ph/0406232v1]}.

\bibitem{semil4}
{X.D Guo, X.Q Hao, H.W Ke, M.G Zhao, X.Qi Li}, ``{Looking for New Physics via
  Semi-leptonic and Leptonic rare decays of $D$ and $D_s$},'' {\em
  DOI:10.1088/1674-1137/41/9/093107}.
\newblock \href{https://arxiv.org/abs/1703.08799v1}{[arXiv:1703.08799v1
  [hep-ph]]}.

\bibitem{Crivellin:2015era}
A.~Crivellin, L.~Hofer, J.~Matias, U.~Nierste, S.~Pokorski, and J.~Rosiek,
  ``{Lepton-flavour violating $B$ decays in generic $Z'$ models},'' {\em Phys.
  Rev.}, vol.~D92, no.~5, p.~054013, 2015.

\bibitem{Crivellin:2016vjc}
A.~Crivellin, G.~D'Ambrosio, M.~Hoferichter, and L.~C. Tunstall, ``{Violation
  of lepton flavor and lepton flavor universality in rare kaon decays},'' {\em
  Phys. Rev.}, vol.~D93, no.~7, p.~074038, 2016.

\bibitem{KKO}
R.Kitano, M.~Koike, and Y.~Okada, ``Detailed calculation of lepton flavor
  violating muon-electron conversion rate for various nuclei,'' {\em
  DOI:10.1103/PhysRevD.66.096002 10.1103/PhysRevD.76.059902}.
\newblock \href{https://arxiv.org/abs/hep-ph/0203110}{[arXiv:hep-ph/0203110]}.

\bibitem{Ilisie}
V.~Ilisie, {\em Concepts in Quantum Field Theory}.
\newblock Springer, 2016, 978-3-319-22966-9.

\bibitem{SMEFT2}
B.~Grzadkowski, M.~Iskrzynski, M.~Misiak, and J.~Rosiek, ``Dimension-six terms
  in the standard model lagrangian,'' {\em JHEP 1010 (2010) 085}.
\newblock \href{https://arxiv.org/abs/1008.4884}{[arXiv:1008.4884 [hep-ph]]}.

\bibitem{SMEFT3}
R.~Alonso, E.~E. Jenkins, A.~V. Manohar, and M.~Trott, ``Renormalization group
  evolution of the standard model dimension six operators iii: Gauge coupling
  dependence and phenomenology,'' {\em JHEP 1404 (2014) 159}.
\newblock \href{https://arxiv.org/abs/1312.2014}{[arXiv:1312.2014 [hep-ph]]}.

\bibitem{SachaKunoCiri}
V.~Cirigliano, S.~Davidson, and Y.~Kuno, ``Spin-dependent µ → e
  conversion,'' {\em Phys. Lett. B 771 (2017) 242
  doi:10.1016/j.physletb.2017.05.053}.
\newblock \href{https://arxiv.org/pdf/1703.02057.pdf}{[arXiv:1703.02057
  [hep-ph]]}.

\bibitem{FJI1}
{S. Bellucci, M. Lusignoli and L. Maiani, Nucl. Phys. B 189 (1981) 329. } {\em
  DOI:10.1016/0550-3213(81)90384-9}.

\bibitem{FJI2}
{G. Buchalla, A. J. Buras and M. K. Harlander, Nucl. Phys. B 337 (1990) 313.}
  {\em DOI:10.1016/0550-3213(90)90275-I}.

\bibitem{futureASSD}
S.~Davidson and A.~Saporta, ``work in progress,''

\bibitem{Kcstdecay}
J.~Rosner, S.~Stone, and R.~S.~V. de~Water, ``Leptonic decays of charged
  pseudoscalar mesons,''
\newblock \href{https://arxiv.org/abs/1509.02220}{[arXiv:1509.02220[hep-ph]]}.

\bibitem{KDform}
{FLAG Working Group}, ``Review of lattice results concerning low-energy
  particle physics,''
\newblock \href{https://arxiv.org/abs/1607.00299}{[arXiv:1607.00299 [hep-ph]]}.

\bibitem{BandDdecaycst}
A.~Bazavov, C.~Bernard, N.~Brown, C.~DeTar, A.~El-Khadra, S.~G. E.~Gámiz,
  U.~Heller, J.~Komijani, A.~Kronfeld, J.~Laiho, P.~Mackenzie, E.~Neil,
  J.~Simone, R.~Sugar, D.~Toussaint, and R.~V. de~Water, ``{B- and D-meson
  leptonic decay constants from four-flavor lattice QCD},''
\newblock \href{https://arxiv.org/abs/1712.09262}{ [arXiv:1712.09262
  [hep-lat]]}.

\bibitem{Bform}
A.~Khodjamirian, T.~Mannel, and N.~Offen, ``Form factors from light-cone sum
  rules with b-meson distribution amplitudes,'' {\em
  DOI:10.1103/PhysRevD.75.054013}.
\newblock
  \href{https://arxiv.org/abs/hep-ph/0611193}{[arXiv:hep-ph/0611193v2]}.

\end{thebibliography}

\end{document}